\documentclass[prD,superscriptaddress,twocolumn,showpacs,nofootinbib,amsmath,amssymb]{revtex4-1}
\usepackage{graphicx,hyperref,color,mathrsfs,bm}
\usepackage[utf8]{inputenc}

\begin{document}
\pagenumbering{arabic}

\title{Effects of a neutrino--dark energy coupling on oscillations of
high-energy neutrinos}

\author{Niki Klop}
\email{l.b.klop@uva.nl}
\affiliation{GRAPPA Institute, University of Amsterdam, 1098 XH
Amsterdam, The Netherlands}
\author{Shin'ichiro Ando}
\affiliation{GRAPPA Institute, University of Amsterdam, 1098 XH
Amsterdam, The Netherlands}
\affiliation{Kavli Institute for the Physics and Mathematics of the
Universe (Kavli IPMU, WPI), Todai Institutes for Advanced Study,
University of Tokyo, Kashiwa, Chiba 277-8583, Japan}
\begin{abstract}

If dark energy (DE) is a dynamical field rather than a cosmological
 constant, an interaction between DE and the neutrino sector could
 exist, modifying the neutrino oscillation phenomenology and causing CP
 and apparent Lorentz violating effects. The terms in the Hamiltonian
 for flavor propagation
 induced by the DE-neutrino coupling do not depend on the neutrino
 energy, while the ordinary
 components decrease as $\Delta m^2/E_{\nu}$. Therefore, the DE-induced
 effects are absent at lower neutrino energies, but become significant
 at higher energies, allowing to be searched for by neutrino
 observatories. We explore the impact of the DE-neutrino coupling on the
 oscillation probability and the flavor transition in the three-flavor
 framework, and investigate the CP-violating and apparent Lorentz
 violating effects. We find that DE-induced effects become observable
 for $E_{\nu}m_{\text{eff}} \sim 10^{-20}~ \text{GeV}^2$, where $m_{\rm
 eff}$ is the effective mass parameter in the DE-induced oscillation
 probability, and CP is
 violated over a wide energy range. We also show that current and future
 experiments have the sensitivity to detect anomalous effects induced by
 a DE-neutrino coupling and probe the new mixing parameters. The
 DE-induced effects on neutrino oscillation can be distinguished from
 other new physics possibilities with similar effects, through the
 detection of the directional dependence of the interaction, which is
 specific to this interaction with DE. However, current experiments will
 not yet be able to measure the small changes of $\sim 0.03\%$ in the
 flavor composition due to this directional effect.
\end{abstract}
\date{\today}
\maketitle

\section{Introduction}
\label{sec:intro}

Dark Energy (DE) is a well established hypothesis in cosmology, being
the driving force behind the accelerated expansion of the Universe. It
makes up for $\sim$68\% of the total energy density in the current
Universe~\cite{Ade:2015xua}.
However, the nature of this presumed DE is still unknown, and several
possible explanations are being considered.
It could be a cosmological constant, which is a constant-valued energy
density through time and space \cite{Weinberg:1988cp, Carroll:2000fy}.
The other possibility is that it is composed of a scalar field, like
quintessence~\cite{Copeland:2006wr, Caldwell:2009ix}.
In the latter case, DE might be able to undergo interactions with
standard model particles, which we can search for in experiments. 

For instance, there could exist a coupling between neutrinos and
dynamical field  DE. 
Such a coupling gives rise to an effective potential, which engenders an
effect on neutrino oscillations that influences the evolution equation
in a way that one could compare with the Mikheyev-Smirnov-Wolfenstein
(MSW) effect that occurs when neutrinos propagate through
matter~\cite{Wolfenstein:1977ue, Mikheev:1986gs,Mikheev:1986wj,
Kuo:1989qe}.
This interaction will change the oscillation probability, and therefore
has an impact on the flavor ratios of the neutrinos detected at Earth. 
The DE-induced part in the Hamiltonian for flavor propagation is
independent of the neutrino energy, while the normal vacuum part falls
off as $1/E_{\nu}$.
Therefore, the effect becomes more significant for higher neutrino
energies, and might be detectable in experiments sensitive to
high-energy extraterrestrial neutrinos such as
IceCube~\cite{Aartsen:2016nxy} and KM3NeT~\cite{Adrian-Martinez:2016fdl}
and ultrahigh energy neutrinos such as ANITA~\cite{Gorham:2008dv} and
Auger~\cite{ThePierreAuger:2015rma}. 
Furthermore, since the expansion of the Universe is going outward in all
directions, the preferred frame of this cosmic expansion is orthogonal
to surfaces of constant DE density. 
Therefore, since we as observers are not in the
cosmic-microwave-background (CMB) rest frame, the effect of the
DE-neutrino interaction does depend on the propagation direction of the
neutrinos. 
This CPT and Lorentz violating coupling has been studied before in
Ref.~\cite{Ando:2009ts}, and in this work, we extend this idea to the
case of thee-neutrino mixing.

With the IceCube detector fully operating and KM3NeT to follow in the
near future, a new window has opened for searches of new physics. 
A general study of new physics through high energetic neutrinos and
their effect on the flavor ratio at Earth is performed in
Ref.~\cite{Arguelles:2015dca}, by introducing effective operators.
(See also Refs.~\cite{Coleman:1998ti, Barger:2000iv, Barenboim:2002rv,
Kostelecky:2003cr, Christian:2004xb, Hooper:2005jp, Katori:2006mz, Kostelecky:2011gq,Diaz:2013wia} for
earlier theoretical work.)
The DE-neutrino coupling that we study is a model that predicts specific
types of terms in the interaction Lagrangian, which engenders such new
physics.
In Refs.~\cite{Bustamante:2015waa, Rasmussen:2017ert} the parameter
space for the flavor ratio at Earth is explored, considering several
beyond-the-standard-model theories that have an impact at the
production, propagation, and detection of astrophysical neutrinos. 
Recently, the IceCube collaboration performed a search for signals of
Lorentz violation in their data of high-energy atmospheric neutrinos
\cite{Aartsen:2017ibm}, and obtained stringent constraints particularly
for higher dimensional operators than the ones that we specifically
study for DE-neutrino couplings. (See Refs.~\cite{GonzalezGarcia:2004wg,
Abbasi:2009nfa, Bahcall:2002ia,Kostelecky:2008ts, Abe:2014wla} for earlier constraints.)
In addition, since, as we show later, the oscillation length of the
DE-induced mixing is much larger than the travel distance of atmospheric
neutrinos, the constraints obtained in \cite{Aartsen:2017ibm} are not
applicable on the DE-neutrino coupling that we study.

In this paper we study the impact of the possible DE-neutrino coupling
on the flavor composition of high-energy extraterrestrial neutrinos and
the consequences of this interaction for current and future
experiments.
We explore the behavior of the probability and the CP violating effects,
as well as the effects of the directional dependence. 
We also determine the sensitivity for experiments to be able to measure
those effects.
 
The paper is organised as follows. 
In Sec.~\ref{sec:theory}, we introduce the theory behind the DE-neutrino
coupling and derive the DE induced oscillation probability in the
framework of three-neutrino mixing. 
Extra details can be found in the Appendix. 
In Sec.~\ref{sec:behaviour}, we explore the effects of the coupling on
the behavior of the oscillations of high-energy neutrinos and discuss
the impact on the flavor composition. 
We also investigate the CP-violating effects. 
In Sec.~\ref{sec:sensitivity}, we determine the sensitivity to those
effects for current and future experiments, and explore the directional
effects in Sec.~\ref{sec:directional}, followed by conclusions in
Sec.~\ref{sec:concl}.

\section{theory}\label{sec:theory}

\subsection{Dark energy--neutrino interaction}

We consider the DE-Neutrino coupling, following the discussion in
Ref.~\cite{Ando:2009ts}. 
Considering three neutrino flavors, the neutrino fields are described by
the Dirac spinor set $\{\nu_e, \nu_\mu, \nu_\tau\}$, and their charge
conjugates by the set $\{\nu_{e^c}, \nu_{\mu^c}, \nu_{\tau^c}\}$. 
The six neutrino fields are combined in the object $\nu_A$, where $A$
runs over the neutrino flavors and their conjugates. 
The most general Lorentz/CPT-violating form of the equations of motion
is then given by~\cite{Kostelecky:2003cr}
\begin{equation}
(i\gamma^{\mu}\delta_{\mu}-M_{AB})\nu_B=0,
\end{equation}
where
\begin{align}\label{eq:massterm}
\begin{split}
M_{AB} \equiv &\,  m_{AB} + im_{5AB}\gamma_5 + a^{\mu}_{AB}\gamma_{\mu}+b^{\mu}_{AB}\gamma_5\gamma_{\mu}\\
+&\frac{1}{2}H^{\mu\nu}_{AB}\sigma_{\mu\nu}.
\end{split}
\end{align}

The four-vectors $a^\mu$, $b^\mu$, and the antisymmetric tensor
$H^{\mu\nu}$ in Eq.~(\ref{eq:massterm}) parametrize Lorentz violation. 
$H^{\mu\nu}$ is only Lorentz violating, while the parameters $a^\mu$ and
$b^\mu$ are CPT violating as well. 
These parameters are highly restricted in our case where the coupling
with DE is responsible for the Lorentz/CPT violation. 
The expansion of the Universe has an outward direction, thus
the unit four-vector that parametrizes the preferred frame of this
cosmic expansion, $l^\mu$, is orthogonal to the surfaces of constant DE
density, which is closely aligned with the surfaces of constant CMB
temperature~\cite{Ando:2009ts,Gordon:2005ai,Erickcek:2008jp,Zibin:2008fe,Erickcek:2008sm}. 
Therefore, $a^\mu \propto l^\mu$ and $b^\mu \propto l^\mu$, where $l^\mu
= (1,0,0,0)$ in the rest frame of the CMB.
Also, $H^{\mu\nu}$ should be proportional to $l^\mu$, but since it is
not possible to create an anti-symmetric tensor from just one
four-vector, $H^{\mu\nu}$ has to be zero in the case of our DE-neutrino
coupling. 
Finally, the DE-neutrino coupling can be parametrized solely by the
combination of $a^\mu$ and $b^\mu$, namely $(a_L)^{\mu}_{ab} \equiv
(a+b)^\mu_{ab}$, where we have $(a_L)^{\mu}_{ab} \propto
l^{\mu}$~\cite{Ando:2009ts, Kostelecky:2003cr}.
Because the velocity of our solar system with respect to the CMB
restframe is $\sim$10$^{-3}$ times the speed of light, we have
$(a_L)^{\mu} p_{\mu} \propto E(1-\bm{v} \cdot \bm{\hat{p}})$, where
$\bm{v}$ is our velocity with respect to the CMB rest frame and
$\bm{\hat{p}}$ is the neutrino propagation direction.

A simple form of Langrangian that describes an interaction by the
DE-neutrino coupling is given by
\begin{equation}\label{eq:lagrangian}
\mathcal{L}_{int} =
 -\lambda_{\alpha\beta}\frac{\partial_{\mu}\phi}{M_{*}}
 \bar{\nu}_{\alpha}\gamma^{\mu}(1-\gamma_5)\nu_{\beta} ,
\end{equation}
where $\phi$ is a quintessence field, $\lambda_{\alpha\beta}$ is a
coupling constant matrix and $M_*$ is the energy scale of the
interaction. 
In this example, we have $a^{\mu}_{L} \sim \lambda
\dot{\phi}(t)l^{\mu}/M_*$. 

The effective Hamiltonian that describes the propagation of the flavor
eigenstates to leading order is given by
\begin{widetext}
\begin{equation}
h_{eff}=
\begin{bmatrix}
p\delta_{ab}+(\tilde{m}^2)_{ab}/2p + (a_L)^{\mu}_{ab}p_{\mu}/p && 0  \\
0 && p\delta_{ab}+(\tilde{m}^2)^*_{ab}/2p - (a_L)^{*\mu}_{ab}p_{\mu}/p
\end{bmatrix},
\end{equation}
\end{widetext}
where the indices $a,b$ run over the flavour eigenstates $e, \mu,
\tau$. 
The upper left block describes the neutrino interactions, and the lower
right the antineutrinos. 
Since the effective Hamiltonian is block diagonal, no mixing will take
place between neutrinos and antineutrinos, and therefore we consider the
two blocks for neutrinos and antineutrinos separately. 

The Hamiltonian that describes the neutrino propagation in vacuum in the
mass base is given by
\begin{equation}
H_{m}=
\begin{bmatrix}
E_1  && 0 && 0 \\
0 && E_2&& 0 \\
0 && 0 && E_3
\end{bmatrix},
\end{equation}
where $E_i = \sqrt{p^2 + m_i^2}$. 
The Hamiltonian in the flavor basis is then obtained by rotating the
basis as
\begin{equation}
H_f = UH_mU^{\dag} ,
\end{equation}
where $U$ is the standard Pontecorvo-Maki-Nakagawa-Sakata (PMNS) matrix
for three-neutrino mixing~\cite{Giganti:2017fhf}.

The Hamiltonian that describes the DE-induced mixing in the basis in
which it demonstrates itself in diagonal form, is given by
\begin{equation}\label{eq:Vm}
V_{m}=
\begin{bmatrix}
\pm k_1(1-\bm{v} \cdot \bm{\hat{p}})  && 0 && 0 \\
0 && \pm k_2(1-\bm{v} \cdot \bm{\hat{p}}) && 0 \\
0 && 0 && \pm k_3(1-\bm{v} \cdot \bm{\hat{p}}) 
\end{bmatrix},
\end{equation}
in which $k_i$ is a constant and the positive (negative) sign is for
neutrinos (antineutrinos), and the Hamiltonian in the flavor basis is
obtained through
\begin{equation}\label{eq:DEham}
V_f = U_{\text{DE}}V_mU^{\dag}_{\text{DE}},
\end{equation}
where $U_{\text{DE}}$ is an independent unitary matrix.

The mixing matrices $U$ and $U_{\text{DE}}$ are parameterized as
\begin{widetext}
\begin{equation}
U_{(\text{DE})} =
\begin{bmatrix}
c_{12}c_{13}&& s_{12}c_{13} && s_{13}e^{-i\delta}\\
-s_{12}c_{23}-s_{13}s_{23}c_{12}e^{i\delta}&&
 c_{12}c_{23}-s_{12}s_{23}s_{13}e^{i\delta} && s_{23}c_{13} \\
s_{12}s_{23}-s_{13}c_{12}c_{23}e^{i\delta} &&
 -s_{23}c_{12}-s_{12}s_{13}c_{23}e^{i\delta} && c_{13}c_{23}
\end{bmatrix}.
\end{equation}
\end{widetext}
In the standard PMNS matrix $U$, $c_{ij}= \cos{\theta_{ij}}$ and
$s_{ij}= \sin{\theta_{ij}}$, where $\theta_{ij}$ are the vacuum mixing
angles and $\delta$ is the CP-violating phase. 
For the values of the vacuum parameters, we use the best fit values from
the Particle Data Group~\cite{Agashe:2014kda}. 
The equivalent mixing matrix for the DE-induced interaction is given by
$U_{\text{DE}}$, where $c_{ij_{\text{DE}}}=
\cos{\theta_{ij_{\text{DE}}}}$ and $s_{ij_{\text{DE}}}=
\sin{\theta_{ij_{\text{DE}}}}$, with $\theta_{ij_{\text{DE}}}$ and
$\delta_{\text{DE}}$ the extra DE-induced mixing angles and CP-violating
phase.

\subsection{Oscillation probabilities}

The Schr\"{o}dinger equation in the flavor basis is given by
\begin{equation}\label{eq:schr}
i\frac{\text{d}}{\text{d}t}\psi_f(t) = \mathscr{H}_f\psi_f(t),
\end{equation}
where
\begin{equation}
\mathscr{H}_f = UH_mU^{\dag} + U_{\text{DE}}V_mU^{\dag}_{\text{DE}} .
\end{equation}
The solution of the Schr\"{o}dinger equation in Eq.~(\ref{eq:schr}) is
\begin{equation}
\psi_f(t)=e^{-i\mathscr{H}_f t}\psi_f(0).
\end{equation}
In order to calculate $U_f(L)\equiv e^{-i\mathscr{H}_f L}$, where we
replaced $t$ with the oscillation distance $L$, we follow
Ref.~\cite{Ohlsson:1999xb}.
A more detailed derivation is summarised in Appendix~\ref{app:Amplitude
of flavour transition}.

The amplitude of the transition from $\nu_\alpha$ to $\nu_\beta$ is
\begin{equation}
A_{\alpha \beta} \equiv \langle \beta | U_f(L)|\alpha \rangle = \phi \sum_{a=1}^3 e^{-iL\lambda_a}M_{a\alpha\beta},
\end{equation}
where $\phi \equiv e^{-iL\text{tr}\mathscr H_f/3}$, $\lambda_a$ are the
eigenvalues of the traceless part of the Hamiltonian $\mathscr H_f$, $T
\equiv \mathscr H_f - (\text{tr}\mathscr H_f) I /3$ and $ M_{a\alpha\beta}$ is defined as
\begin{equation}
 M_{a\alpha\beta}\equiv\frac{(\lambda_a^2+c_1)\delta_{\alpha\beta} +
  \lambda_a T_{\alpha\beta} + (T^2)_{\alpha\beta}}{3\lambda_a^2+c_1},
\end{equation}
where $c_1 =
T_{11}T_{22}-T_{12}T_{21}+T_{11}T_{33}-T_{13}T_{31}+T_{22}T_{33}-T_{23}T_{32}$.
We can calculate the oscillation probability with
 \begin{equation}
 P_{\alpha \rightarrow \beta} \equiv |A_{\alpha \beta} |^2 .
\end{equation}

Since $T$ is Hermitian ($T^\dagger = T$), the three eigenvalues
$\lambda_a$ are all real.
We now define
\begin{eqnarray}
 c_a &=& \cos(L\lambda_a),\\
 s_a &=& \sin(L\lambda_a),\\
 \mathcal R_{a\alpha\beta} &=& \mbox{Re}[M_{a\alpha\beta}],\\
 \mathcal I_{a\alpha\beta} &=& \mbox{Im}[M_{a\alpha\beta}],
\end{eqnarray}
and rewrite the oscillation probability as
\begin{eqnarray}
 P_{\alpha\beta} 
  &=&\sum_{ab}\left[(c_ac_b+s_as_b)(\mathcal R_{a\alpha\beta}\mathcal
	    R_{b\alpha\beta}+\mathcal I_{a\alpha\beta}\mathcal
	    I_{b\alpha\beta})
  \right.\nonumber\\&&{}\left.
  +(s_ac_b-s_bc_a)(\mathcal R_{b\alpha\beta}\mathcal
  I_{a\alpha\beta} -R_{a\alpha\beta}\mathcal
  I_{b\alpha\beta}) \right].
\end{eqnarray}
We further use
\begin{eqnarray}
 c_ac_b+s_as_b &=& 1-2\sin^2x_{ab}\\
 s_ac_b-s_bc_a &=& 2\sin x_{ab}\cos x_{ab},
\end{eqnarray}
where $x_{ab} = (\lambda_a-\lambda_b)L/2$, and arrive at
\begin{widetext}
\begin{equation}\label{eq:probab}
 P_{\alpha\beta} = \delta_{\alpha\beta}-4\sum_a\sum_{b<a}
  \left[(\mathcal R_{a\alpha\beta}\mathcal R_{b\alpha\beta} + \mathcal
   I_{a\alpha\beta}\mathcal I_{b\alpha\beta})\sin^2x_{ab}\right]
    +2\sum_a\sum_{b<a}\left[(\mathcal R_{b\alpha\beta}\mathcal
  I_{a\alpha\beta} -R_{a\alpha\beta}\mathcal
  I_{b\alpha\beta})\sin 2x_{ab}\right].
\end{equation}
\end{widetext}
Here in obtaining the first term, we used the fact that
$P_{\alpha\beta} = \delta_{\alpha\beta}$ at $L = 0$.

Rather than on the individual parameters $k_i$ that show up in the
Hamiltonian of Eq.~(\ref{eq:Vm}), the probability will depend on the
differences $k_j-k_i$, which we call the effective mass parameter,
$m_{\text{eff}_{ji}} \equiv k_j - k_i$.
Since we consider the three-flavor case, two of them are
independent: $m_{\text{eff}_{21}} \equiv k_2 - k_1$ and
$m_{\text{eff}_{31}} \equiv k_3 - k_1$. 
When both independent effective mass parameters equal to zero,
Eq.~(\ref{eq:probab}) returns the vacuum oscillation probability.

For distances much larger than the oscillation length, we may replace
$\sin^2x_{ab} \to 1/2$ and $\sin 2 x_{ab} \to 0$, while for distances
much shorter than the oscillation length, it is not possible to observe
effects induced by the DE-neutrino coupling. For example,  if the
effective mass parameter has a value of $m_{\rm eff }=
10^{-23}~\text{GeV}$, the oscillation length is approximately
$L_{\text{osc}} \sim 10^{14}$~km.
Since in our case, we are interested in astrophysical neutrinos, the
probability that we use reduces to
\begin{equation}\label{eq:proba}
 P_{\alpha\beta} = \delta_{\alpha\beta}-2\sum_a\sum_{b<a}
  \left[(\mathcal R_{a\alpha\beta}\mathcal R_{b\alpha\beta} + \mathcal
   I_{a\alpha\beta}\mathcal I_{b\alpha\beta})\right].
\end{equation}
This is justified especially for sources at cosmological distances, $L
\sim H_0^{-1}$, which is equivalent to assuming $m_{\rm eff} \gg H_0
\approx 10^{-42}$~GeV.
In the next section, we shall see that this is indeed the case for the
values of $m_{\text{eff}}$ that we consider.

As can be seen from the DE-induced Hamiltonian in Eq.~(\ref{eq:DEham}),
the DE-induced part of the probability has different sign for neutrinos
and anti-neutrinos; i.e., CP is violated. 
It also does not depend on the neutrino energy, while the vacuum
probability falls off over $E_\nu$. 
Therefore, the impact of DE on neutrino oscillations will become more
significant for higher neutrino energies, and thus the effect could be
explored through experiments such as IceCube and KM3NeT. 

Finally, the DE-induced part is frame dependent. 
It depends on our velocity with respect to the CMB rest frame, and the
propagation direction of the incoming neutrino.

To summarize, the probability will depend on three new mixing angles,
one extra CP-violating phase, and two independent effective mass
parameters. 
We will investigate the impact of the DE-neutrino coupling on neutrino
oscillations and explore how the probability behaves for different values of the
new mixing parameters in the next section. 
Throughout this work, we assume normal mass hierarchy.

\section{Results}

\subsection{Behavior of the probability}\label{sec:behaviour}

\begin{figure}
\includegraphics[width=0.55\textwidth]{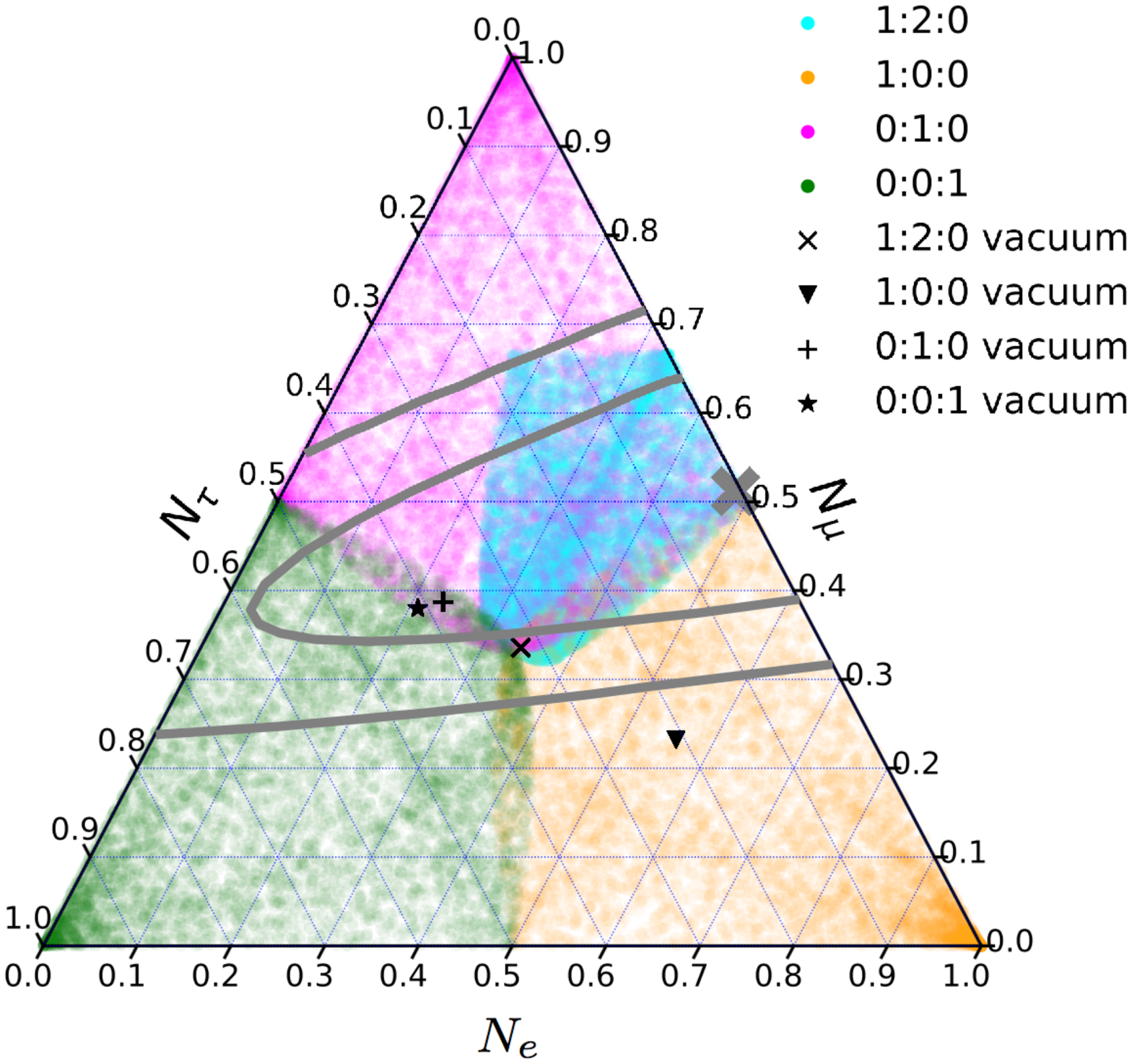} 
\caption{The possible ratios of $\nu_e$:$\nu_\mu$:$\nu_\tau$ at Earth
 for different starting flavor ratios $\nu_e$:$\nu_\mu$:$\nu_\tau$ at
 the source. The colored regions correspond to oscillation in the
 presence of DE-induced mixing, where we varied over all combinations of
 the values of the new mixing angles. The expected ratios for vacuum
 mixing (assuming normal hierarchy) are drawn in
 black. The solid grey contours show the allowed regions by IceCube at 68\% and
 95\% confidence levels, while the grey cross represents their best-fit flavor
 ratio \cite{Aartsen:2015knd}.}\label{fig:triangle}
\end{figure}

\begin{figure}
\includegraphics[width=0.55\textwidth]{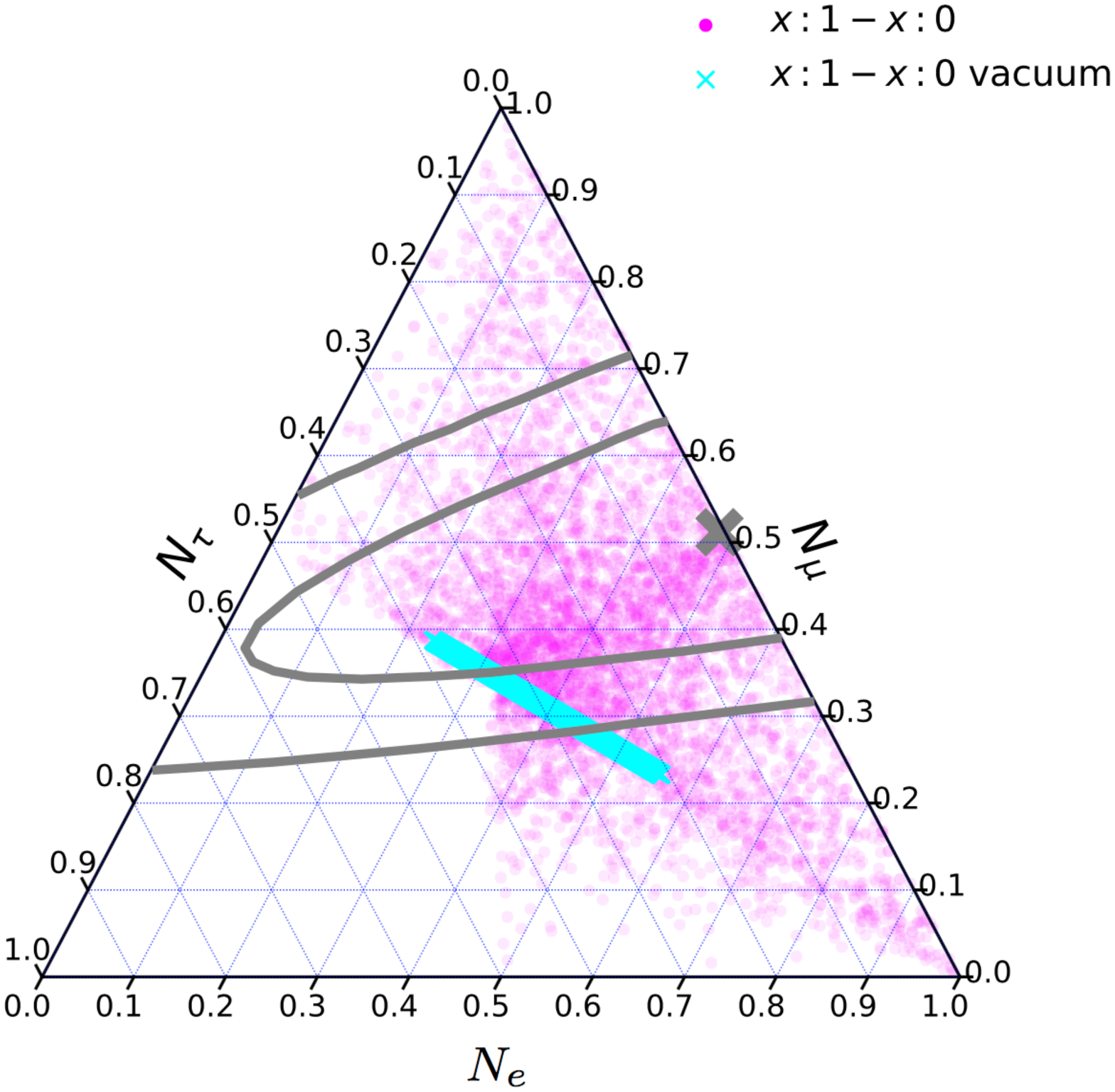} 
\caption{The possible ratios of  $\nu_e$:$\nu_\mu$:$\nu_\tau$ at Earth
 for different proportions of $\nu_e$ and $\nu_{\mu}$ at the source, and
 no $\nu_{\tau}$ at production. The magenta region corresponds to
 oscillation in the presence of DE-induced mixing, where all
 combinations of the values of the new mixing angles are varied. The
 cyan region corresponds to vacuum oscillation. The solid grey contours show the allowed regions by IceCube at 68\% and
 95\% confidence levels, while the grey cross represents their best-fit flavor
 ratio \cite{Aartsen:2015knd}.}\label{fig:combi}
\end{figure}

\begin{figure*}[t!]
\includegraphics[trim={0 6.5cm 0 6.5cm},width=0.47\textwidth]{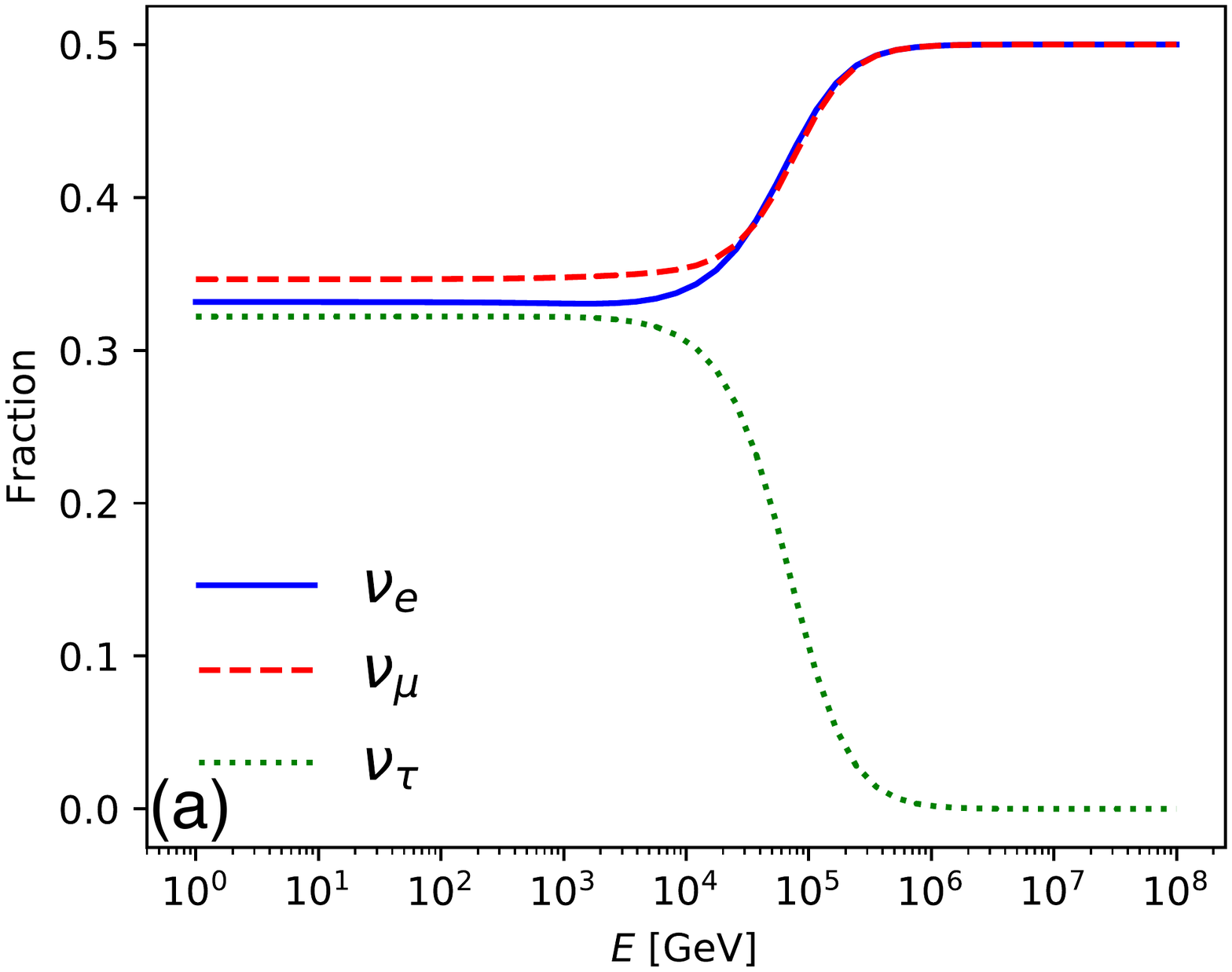} 
\includegraphics[trim={0 6.5cm 0 6.5cm},width=0.47\textwidth]{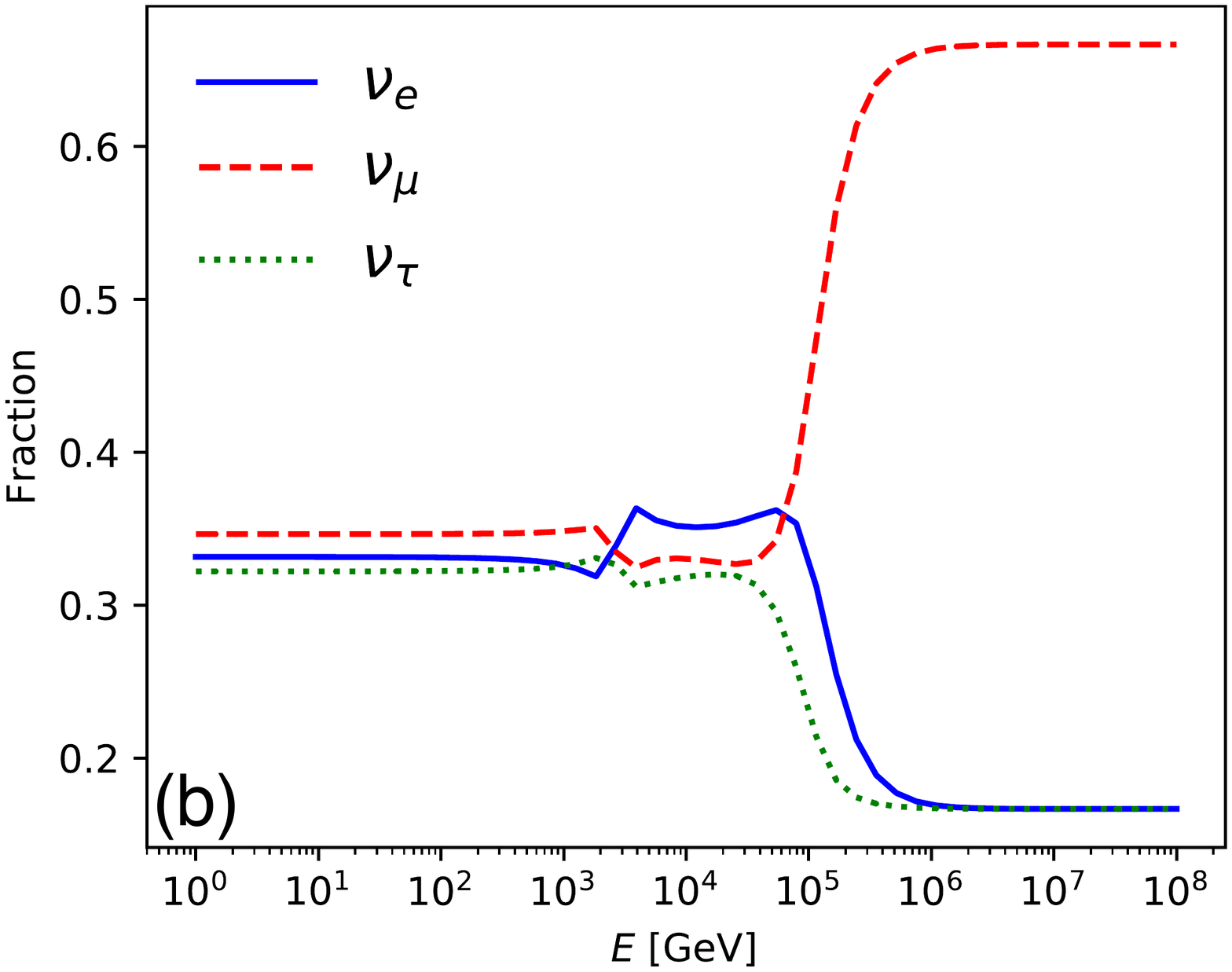}
\includegraphics[trim={0 6.5cm 0 6.5cm},width=0.47\textwidth]{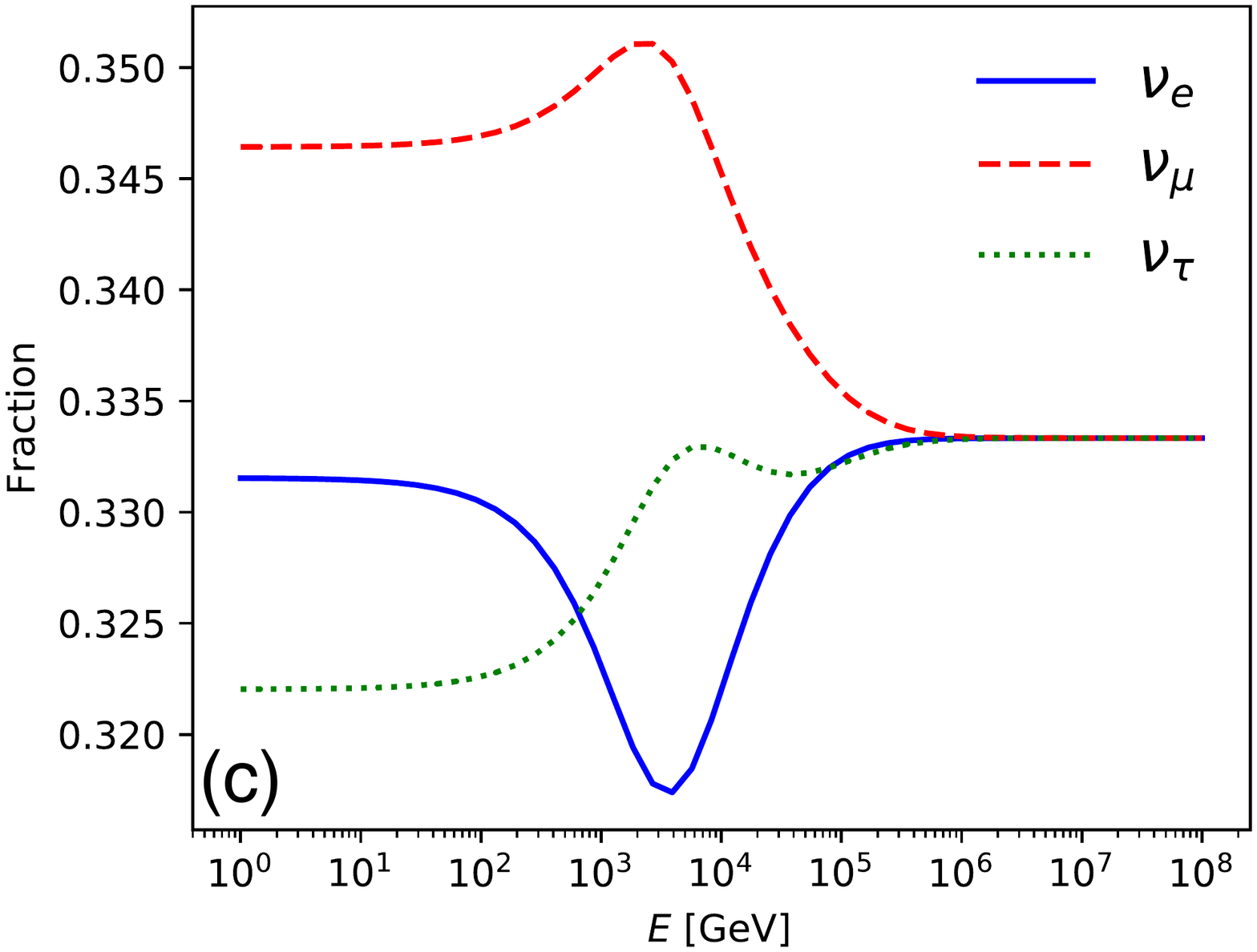}
\includegraphics[trim={0 6.5cm 0 6.5cm},width=0.47\textwidth]{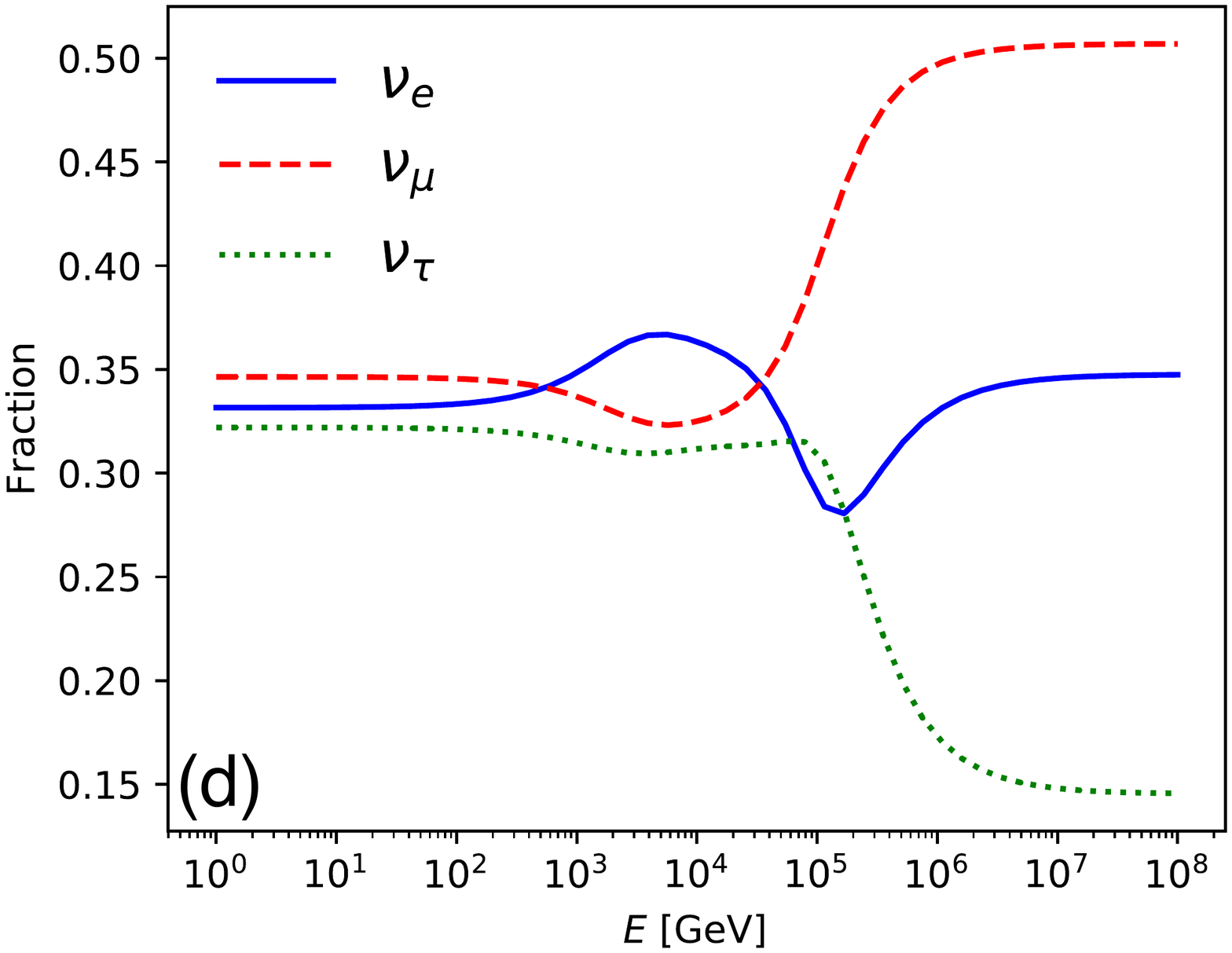}
\caption{The flavour ratios as a function of neutrino energy for
 different sets of parameter values. The flavor composition at the
 source is set to 1:2:0. The effective mass parameters are set to
 $m_{\text{eff}_{21}} = \frac{1}{2}m_{\text{eff}_{31}} = 10^{-26}
 \text{GeV}$, and $\delta_{\text{CP}}=0$. The values of the new mixing
 angles are set to (a) $\theta_{12}= 0.25 \pi$ and
 $\theta_{13},\theta_{23} = 0$; (b) $\theta_{13}= 0.25 \pi$ and
 $\theta_{12},\theta_{23} = 0$; (c) $\theta_{23}= 0.25 \pi$ and
 $\theta_{12},\theta_{13} = 0$ (bottom left); and (d)
 $\theta_{12},\theta_{13},\theta_{23} = 0.25 \pi$ (maximal mixing). The
 transition from the domination of vacuum oscillation to DE-induced
 domination takes place at $E_{\nu}m_{\text{eff}} \sim 10^{-20}$
 GeV$^2$.}
\label{fig:fractions}
\end{figure*}

\begin{figure}[t!]
\includegraphics[trim={0 6.5cm 0 6.5cm},clip,width=0.43\textwidth]{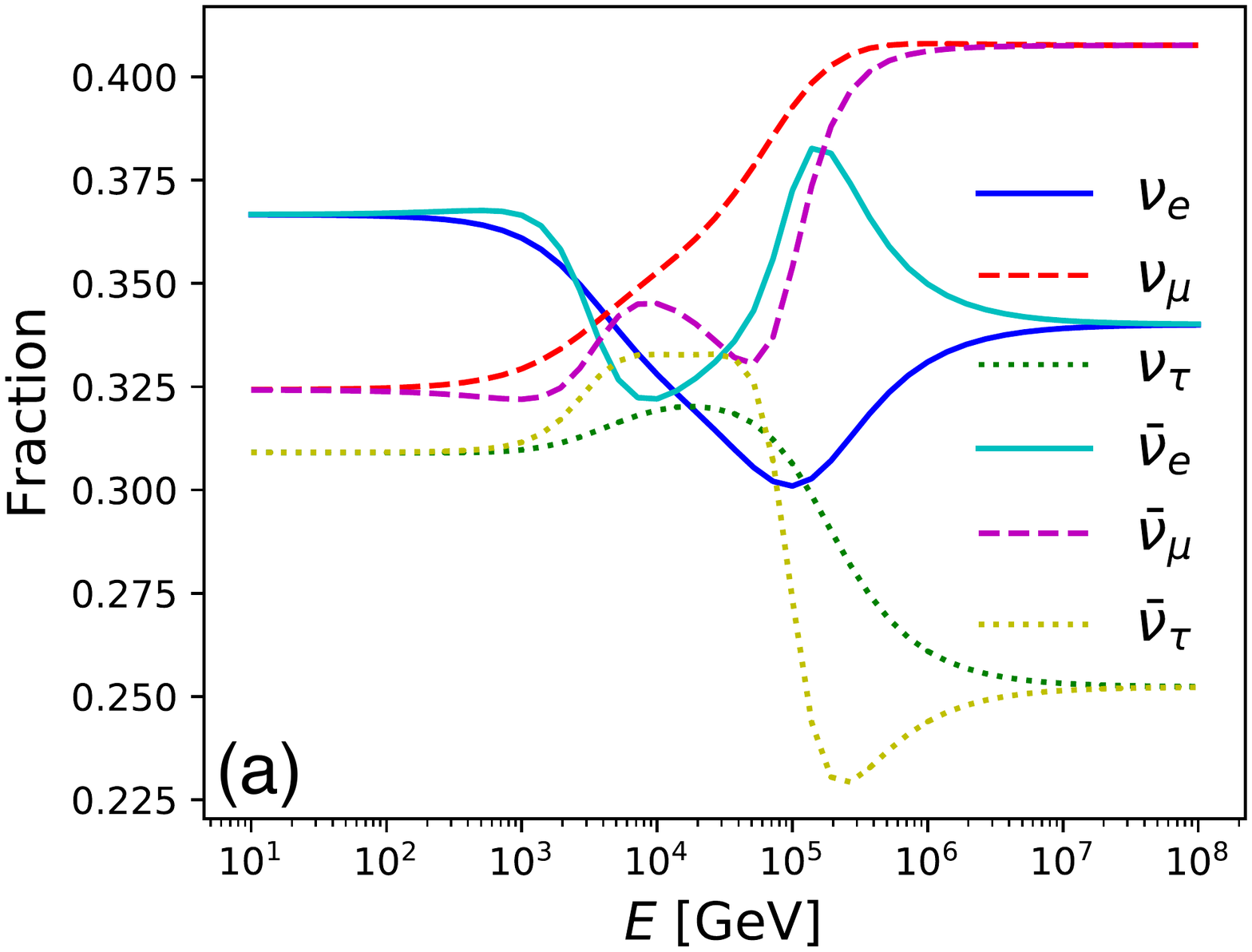} 
\includegraphics[trim={0 6.5cm 0 6.5cm},clip,width=0.43\textwidth]{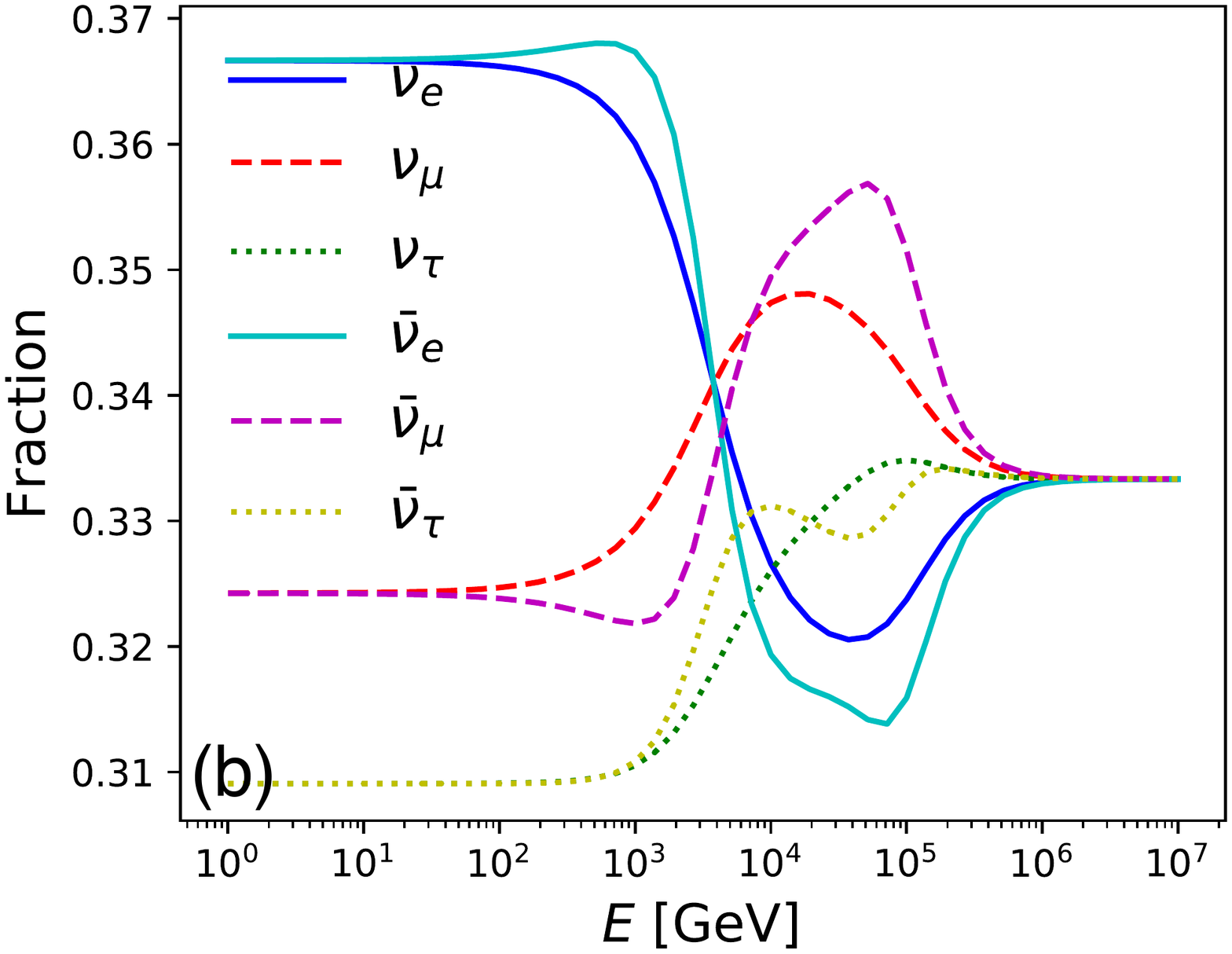}
\includegraphics[trim={0 6.5cm 0 6.5cm},clip,width=0.43\textwidth]{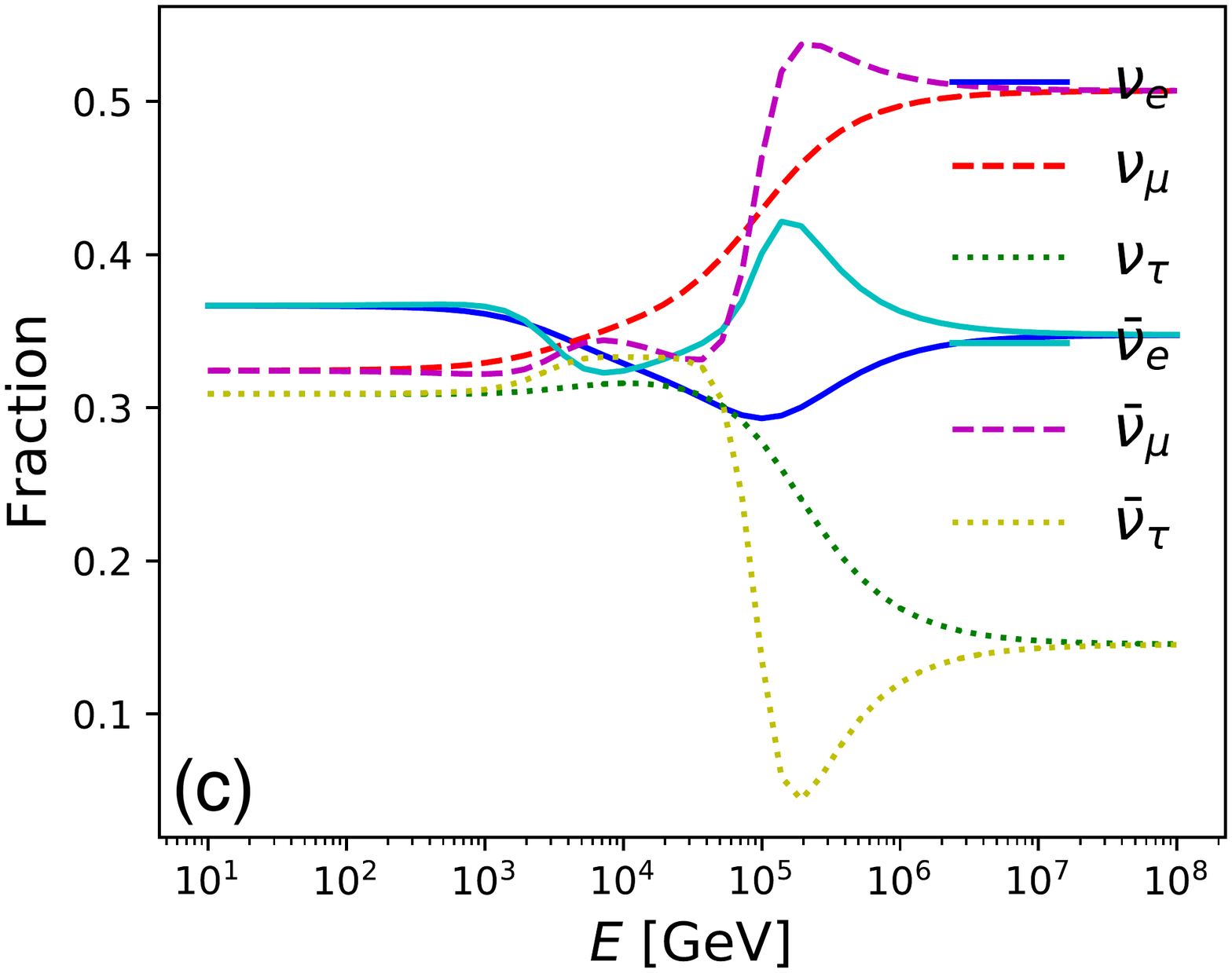}
\caption{The flavour ratios as a function of neutrino energy for
 different values of $\delta_{\text{DE}}$, for both neutrinos and
 anti-neutrinos. The values of the new mixing parameters are set to
 $\theta_{12},\theta_{13},\theta_{23} = 0.25 \pi$, $m_{\text{eff}_{21}}
 = \frac{1}{2}m_{\text{eff}_{31}} = 10^{-26} \text{GeV}$ and the starting flavor
 ratio is set to 1:2:0. The usual CP-violating phase
 $\delta_{\text{DE}}$ is set to zero. The new DE-induced CP-violating
 phase is set to (a) $0.25 \pi$, (b) $0.5\pi$, and (c) 0. The effect in
 (c) comes solely from the sign difference of $m_{\text{eff}}$ in the
 oscillation probability for neutrinos and
 anti-neutrinos.}
\label{fig:CPV} 
\end{figure}

To explore the effect of the DE-neutrino coupling on what we detect here
at Earth, we determined the possible final flavor compositions at the
time of detection in the presence of this coupling. 
The result can be seen in Fig.~\ref{fig:triangle}. 
We varied all the values of the new mixing parameters, and determined
the final flavor composition for several starting flavor ratios at the
source. 
The expected composition for vacuum oscillation is also included, for
which the mixing parameters are fixed at the best-fit values of the
Particle Data Group~\cite{Agashe:2014kda}. 

As can be seen, the part of the composition-triangle that could be
reached at Earth, depends on the flavor composition at the source. 
The cyan colored area corresponds to the source composition 1:2:0 for
the flavors $e$:$\mu$:$\tau$, which is the characteristic flavor
composition from pion decays. 
This is the main channel in which astrophysical neutrinos are expected
to be produced. 
In the case that there is no new physics, the expected flavor
composition measured at detection is approximately 1:1:1 as shown as the
``cross'' symbol. 

Starting from a purely single flavor state, the possible area after
these DE-neutrino interactions can occupy only one-third of the entire
triangle. 
No astrophysical process is known to produce $\tau$ neutrinos. 
In Fig.~\ref{fig:combi}, the possible flavor compositions are shown for
all flavor compositions at the source consisting of a combination of
$\nu_e$ and $\nu_\mu$. 
The cyan colored region corresponds to the case that there is no new
physics. 
If the observed flavor composition lies outside the cyan region, then it
is not compatible with normal oscillation, and regarded as an indication
of new physics. 
If the ratio lies in the magenta region, this could be due to the
DE-neutrino coupling. 
The lower left part of the triangle cannot be reached by conventional
astrophysical neutrinos even with an effect of the DE-neutrino coupling
we study. 
Therefore, it requires both $\nu_\tau$ production at the source and
non-standard neutrino oscillation such as the DE-neutrino interaction
(Fig.~\ref{fig:triangle}).

In Fig.~\ref{fig:fractions}, we explore the behavior of the probability
as a function of energy. 
In these plots, the flavor composition at the source is set to 1:2:0,
and the values of the two independent effective mass parameters are set
to $m_{\text{eff}_{21}} = m_{\text{eff}_{31}}/2 = 10^{-26}$ GeV. 
We set the new CP-violating phase equal to zero, and consider the cases
that $\theta_{\text{DE}_{12}}=0.25 \pi$, $\theta_{\text{DE}_{13}},
\theta_{\text{DE}_{23}}=0$ (Fig.~\ref{fig:fractions}a),
$\theta_{\text{DE}_{13}}=0.25 \pi$, $\theta_{\text{DE}_{12}},
\theta_{\text{DE}_{23}}=0$ (Fig.~\ref{fig:fractions}b),
$\theta_{\text{DE}_{23}}=0.25 \pi$, $\theta_{\text{DE}_{12}},
\theta_{\text{DE}_{13}}=0$ (Fig.~\ref{fig:fractions}c) and
$\theta_{\text{DE}_{12}}=\theta_{\text{DE}_{13}}=\theta_{\text{DE}_{23}}
=0.25 \pi$ (maximal mixing, Fig.~\ref{fig:fractions}d). 
As visible from the plots, for lower energies, vacuum oscillation is
still dominant. 
After a transition phase, that happens around $E_{\nu}m_{\text{eff}}
\sim 10^{-20}$~ GeV$^2$, the mixing caused by the DE-neutrino coupling
dominates.

We also explore the CP-violating effect of the DE-neutrino coupling. 
Figure~\ref{fig:CPV} shows that neutrinos mix differently from
antineutrinos.
In Fig.~\ref{fig:CPV}a, all new mixing angles are set to maximal,
$0.25\pi$, and the new CP-violating phase is also set to maximal.
The CP-violating effect is visible over a wide energy range.
Although IceCube and KM3NeT cannot distinguish between neutrinos and
anti-neutrinos in general, they can recognize $\bar\nu_e$ through the
Glashow resonance~\cite{PhysRev.118.316} by the measurement of the
$W^-$-boson produced on-shell ($\bar\nu_e e^- \to W^-$). 
Thus, if the Glashow resonance-energy of 6.3~PeV lies in the energy
range of the CP-violating effect, it would be possible to distinguish
electron neutrinos from electron antineutrinos at this energy, and
therefore to detect the CP-violating effect. 
In Figs.~\ref{fig:CPV}b and \ref{fig:CPV}c, the new CP-violating phase
is fixed to $\delta_{cp}=0.5$ and $\delta_{cp}=0$, respectively.
The case that $\delta_{cp}=0$ in Fig.~\ref{fig:CPV}c is interesting,
because the CP-violation is not induced by the new CP-violating phase,
but is entirely due to the sign difference between neutrinos and
anti-neutrinos in the Hamiltonian in Eq.~(\ref{eq:Vm}). 
In the case of Fig.~\ref{fig:CPV}b, it is interesting to note that the
flavor composition at earth changes from approximately 1:1:1, to exactly
1:1:1. 
This is also the case in Fig.~\ref{fig:fractions}c.

\subsection{Sensitivity}\label{sec:sensitivity} 

\begin{figure}[t!]
\includegraphics[trim={0 6.5cm 0 6.5cm},width=0.47\textwidth]{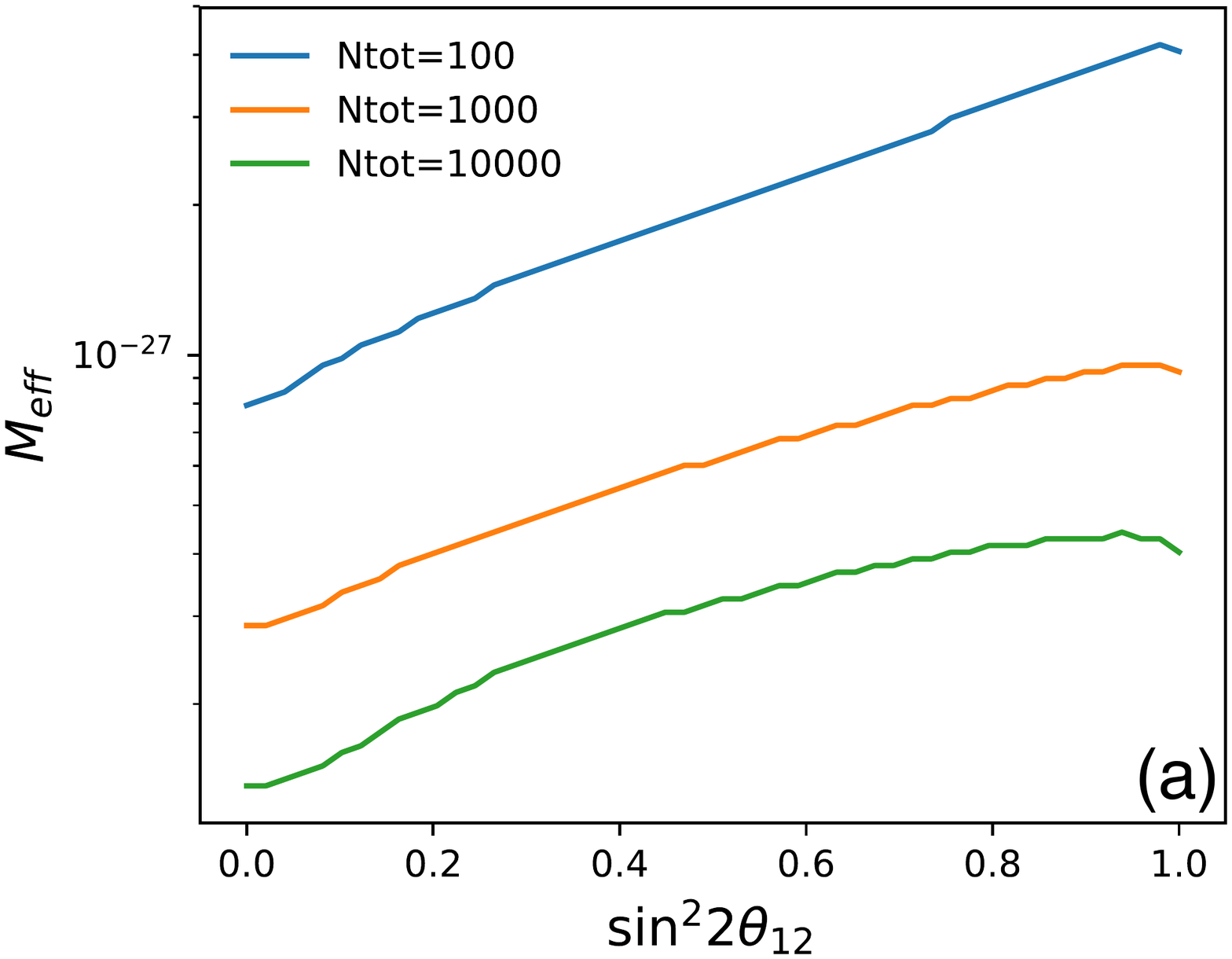} 
\includegraphics[trim={0 6.5cm 0 6.5cm},width=0.47\textwidth]{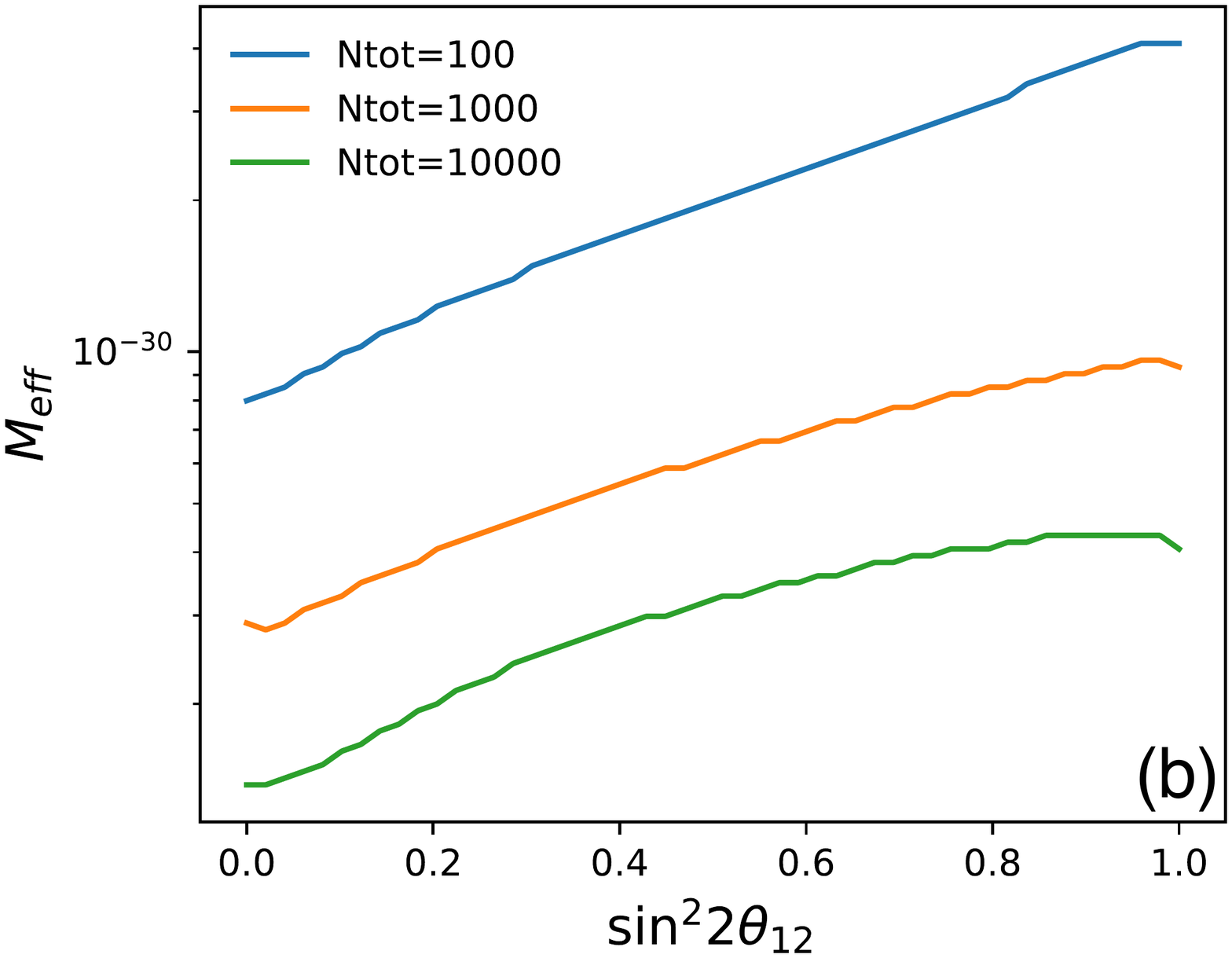}
\caption{The sensitivity for neutrino experiments capable of
 distinguishing muon neutrinos to probe for the value of
 $m_{\text{eff}}$, in case of a detection of 100, 1000 and 10000 events,
 for different neutrino energy ranges: (a) 0.1--100~PeV and (b)
 0.1--100~EeV. When the genuine values of the parameters
 $m_{\text{eff}}$ and $\theta_{12}$ lie above the colored lines, more
 $\nu_{\mu}$ than compatible with standard physics will be detected. In
 this example, the new mixing angles are set to
 $\theta_{13},\theta_{23}=0$. The flavor composition at the source is
 set to 1:2:0.}
\label{fig:sensitivity}
\end{figure}

Fig.~\ref{fig:triangle} shows that all the possible final flavor compositions from the initial
flavor ratio of 1:2:0 are still allowed in light of the IceCube constraints \cite{Aartsen:2015knd}.
To this end, we also investigated the sensitivity for experiments to be
able to detect the effects from the DE-neutrino coupling. 
For this, we compare the total amount of muon neutrinos with the null
hypothesis that no new physics is detected. 
We calculate this for the case that $\theta_{13_{\rm DE}},
\theta_{23_{\rm DE}}=0$, corresponding to the case explored in
Fig.~\ref{fig:fractions}a. 
We set limits on the parameter space for the effective mass parameter
$m_{\text{eff}}$ and the mixing angle $\theta_{12_{\rm DE}}$.
If a number of $N_{\nu}^{\text{tot}}$ neutrino events is detected at the
experiment, assuming a flavor ratio of 1:2:0 at the source, the number
of $\nu_{\mu}$ we expect to measure is given by
\begin{eqnarray}
N_{\nu_{\mu}} &=& \frac{N^{\text{tot}}_{\nu}E_{\text{min}}}{3}\int^{E_{\text{max}}}_{E_{\text{min}}}E^{-2}P_{e\mu}\mathrm{d}E \nonumber\\
&&{}+ \frac{2N^{\text{tot}}_{\nu}E_{\text{min}}}{3}\int^{E_{\text{max}}}_{E_{\text{min}}}E^{-2}P_{\mu\mu}\mathrm{d}E .
\end{eqnarray}
Here we assume that the neutrino energy spectrum multiplied by the effective area
roughly scales as $E^{-2}$.
In case that no new physics is detected, the expected number of muon
neutrinos is $N_{\nu_{\mu}} = N^{\text{tot}}_{\nu}/3$. 
To obtain the limits on $m_{\text{eff}}$ and $\theta_{12_{\rm DE}}$ with
$95\%$ confidence level, we solve
\begin{equation}
N_{\nu_{\mu}} < \frac{N_{\nu}^{\text{tot}}}{3} +
 2\sqrt{\frac{N_{\nu}^{\text{tot}}}{3}}
\end{equation}
for $m_{\text{eff}}$ and $\theta_{12_{\rm DE}}$, where we choose
$m_{\text{eff}_{31}} = 2 m_{\text{eff}_{21}}$. 
In Fig.~\ref{fig:sensitivity}a, we show the sensitivity to probe for the
value of $m_{\text{eff}}$ for experiments measuring neutrino events in
the energy range from 100~TeV to 10~PeV --- which holds for, for
example, IceCube and KM3NeT --- in case that they measure 100, 1000 and
10000 neutrino events, as a function of the mixing angle
$\theta_{12_{\rm DE}}$. 
Given that IceCube already has found tens of neutrino events above
$\sim$10~TeV~\cite{Aartsen:2016xlq, IceCubeICRC2017}, proper analysis
will enable to exclude $m_{\rm eff}\agt 10^{-27}$~GeV in the near
future.
In Fig.~\ref{fig:sensitivity}b, we show the same for (future)
experiments sensitive to ultrahigh-energy (UHE) neutrinos, capable of
detecting neutrinos in the energy range between 100~PeV and 10~EeV. 
Values of $m_{\text{eff}}$ and the corresponding values of
$\theta_{12_{\rm DE}}$ that lie above the coloured curves would result
in an atypical increase of the amount of $\nu_{\mu}$ at the detector. 
In a similar way, this could be calculated for $\nu_{e}$ and
$\nu_{\tau}$.

Looking back at the toy model for a possible Lagrangian of the
DE-neutrino coupling could look like in Eq.~(\ref{eq:lagrangian}), we
follow Ref.~\cite{Ando:2009ts} to explore the mass scale of the
interaction corresponding to a certain value of the effective mass
parameter $m_{\text{eff}}$. 
We have $a^{\mu}_{L} \sim \lambda \dot{\phi}(t)l^{\mu}/M_*$ and $m_{eff}
\sim \Delta\lambda \dot{\phi}(t)/M_*$, with $\Delta\lambda$ the
difference between the eigenvalues of $\lambda_{\alpha\beta}$. 
For quintessence we assume $\dot{\phi} \sim
M_{\text{Pl}}H_{0}(1+w)^{1/2}$~\cite{Caldwell:2009ix,Ando:2009ts}, where
$M_{\text{Pl}}$ is the Planck mass. 
The energy scale of the interaction is therefore given by
\begin{equation}
M_* \simeq 10^6 (\Delta\lambda)\left(\frac{1+w}{0.01} \right)^{1/2}\left(\frac{10^{-30}\text{GeV}}{m_{\text{eff}}} \right) \text{GeV}.
\end{equation}
Therefore, the experiments corresponding to Fig.~\ref{fig:sensitivity}a
and Fig.~\ref{fig:sensitivity}b probe up to mass scales of $M_* \sim
10^5$ GeV and $M_* \sim 10^8$ GeV respectively.

\subsection{Directional dependence}\label{sec:directional}

There are multiple new physics hypotheses that could result in a flavor
composition that is not compatible with normal physics (see, e.g.,
Refs.~\cite{Arguelles:2015dca, Bustamante:2015waa, Rasmussen:2017ert}),
and the DE-neutrino coupling is just one of the possibilities.  
However, the directional dependence of the DE-neutrino coupling is very
specific to this model. 
The DE-induced part of the probability is proportional to $(1-\bm{v}
\cdot \bm{\hat{p}})$, and therefore results in a different mixing
probability for identical neutrinos with different propagation
directions. 
Since our velocity with respect to the CMB rest frame is $\sim 10^{-3}
c$, the effects of the directional component will be small compared to
the general effects of the DE-neutrino coupling. 
To calculate the directional effect, we follow Ref.~\cite{Ando:2009ts}
with some modifications to set our coordinate system. 
The origin is set at the south pole of the Earth, with the $z$-axis
aligned along the rotational axis of the Earth, such that the north pole
lies on the positive axis. 
The $x$-axis is set along the direction to the Sun at spring equinox,
while the $y$-axis is set along this direction at summer solstice. 
The seasonal rotation can be expressed by the azimuthal angle $\phi_s$,
where $\phi_s = 0$ and $\phi_s=\pi$ for spring and autumn equinox
respectively.
Since the velocity of the Sun with respect to the CMB rest frame is
$v_\odot = 369$ km s$^{-1}$ in the direction $\alpha= 168^{\circ}$,
$\delta = -7.22^\circ$, where $\alpha$ and $\delta$ are right ascension
and declination respectively, in our coordinate system this velocity is
$\bm{v}_\odot = v_\odot
(\cos{\delta}\cos{\alpha},\cos{\delta}\sin{\alpha},\sin{\delta}) =
(-385, 76.1, -46.4)$ km s$^{-1}$. 
Because the Earth moves around the Sun with an average orbital speed of
$v_\oplus = 29.8$ km s$^{-1}$, the velocity of the Earth with respect to
the CMB rest frame is
\begin{eqnarray}
\bm{v}_\oplus & = & \bm{v}_\odot + v_\oplus\begin{pmatrix} \sin{\phi_s} \\ -\cos{\phi_s}\cos{\theta_{\text{inc}}}\\  -\cos{\phi_s}\sin{\theta_{\text{inc}}} \end{pmatrix} \nonumber\\
& = &\begin{pmatrix} -358 +29.8\sin{\phi_s} \\ 76.1 -27.3\cos{\phi_s}\\  -46.4 -11.9\cos{\phi_s} \end{pmatrix} \text{km s}^{-1},
\end{eqnarray}
where $\theta_{\text{inc}}$ is the inclination between the $x$-$y$ plane
and the orbital plane around the Sun. 
In our coordinate frame, the south pole is set to $(0,0,0)$. The
propagation direction of the incoming neutrino is therefore described by
the unit vector
\begin{equation}
\bm{\hat{p}} = \begin{pmatrix} \cos{\theta_{\nu}}\cos{\phi_{\nu}} \\ \sin{\theta_{\nu}}\cos{\phi_{\nu}} \\ \sin{\phi_{\nu}}  \end{pmatrix},
\end{equation}
where $\phi_\nu$ and $\theta_\nu$ are the polar and azimuthal angle of
the incoming neutrino at the south pole respectively. 
Since the source lies outside Earth, the propagation direction of the
neutrino path with respect to the CMB background does not depend on the
rotation of the Earth, although it would in the case of an Earth-based
neutrino beam. 
The mixing probability has terms that are proportional to $(1-\bm{v}
\cdot \bm{\hat{p}})^2$,  $(1-\bm{v} \cdot \bm{\hat{p}})$ and terms that
are not dependent on $(1-\bm{v} \cdot \bm{\hat{p}})$ at all.
We expect the effect to be larger when the terms $\propto (1-\bm{v}
\cdot \bm{\hat{p}})$ dominate, which is the case in the transition
phase.

\begin{figure*}
\includegraphics[trim={0 7.5cm 0 7.5cm},clip,width=0.47\textwidth]{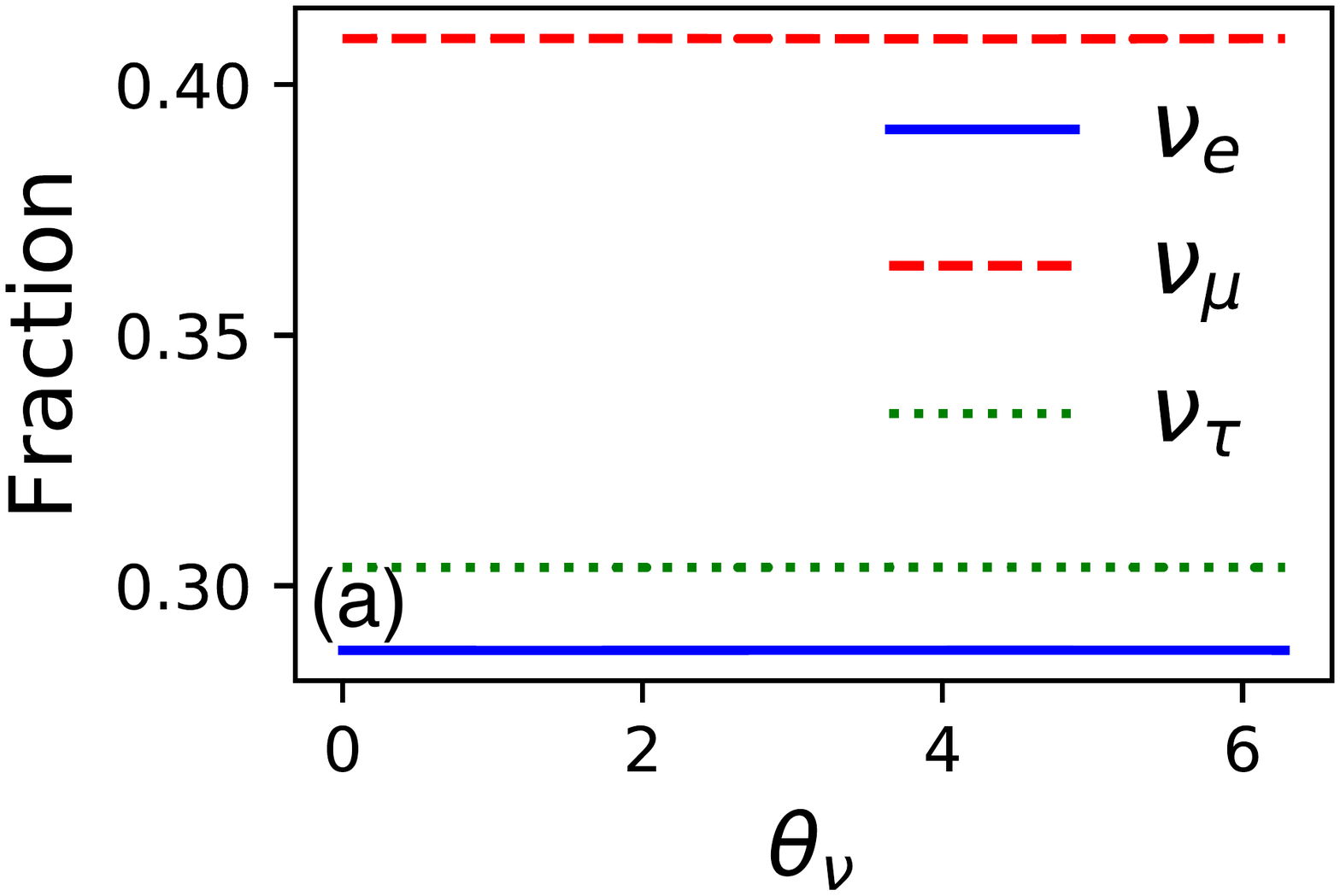} 
\includegraphics[trim={0 7.5cm 0 7.5cm},clip,width=0.47\textwidth]{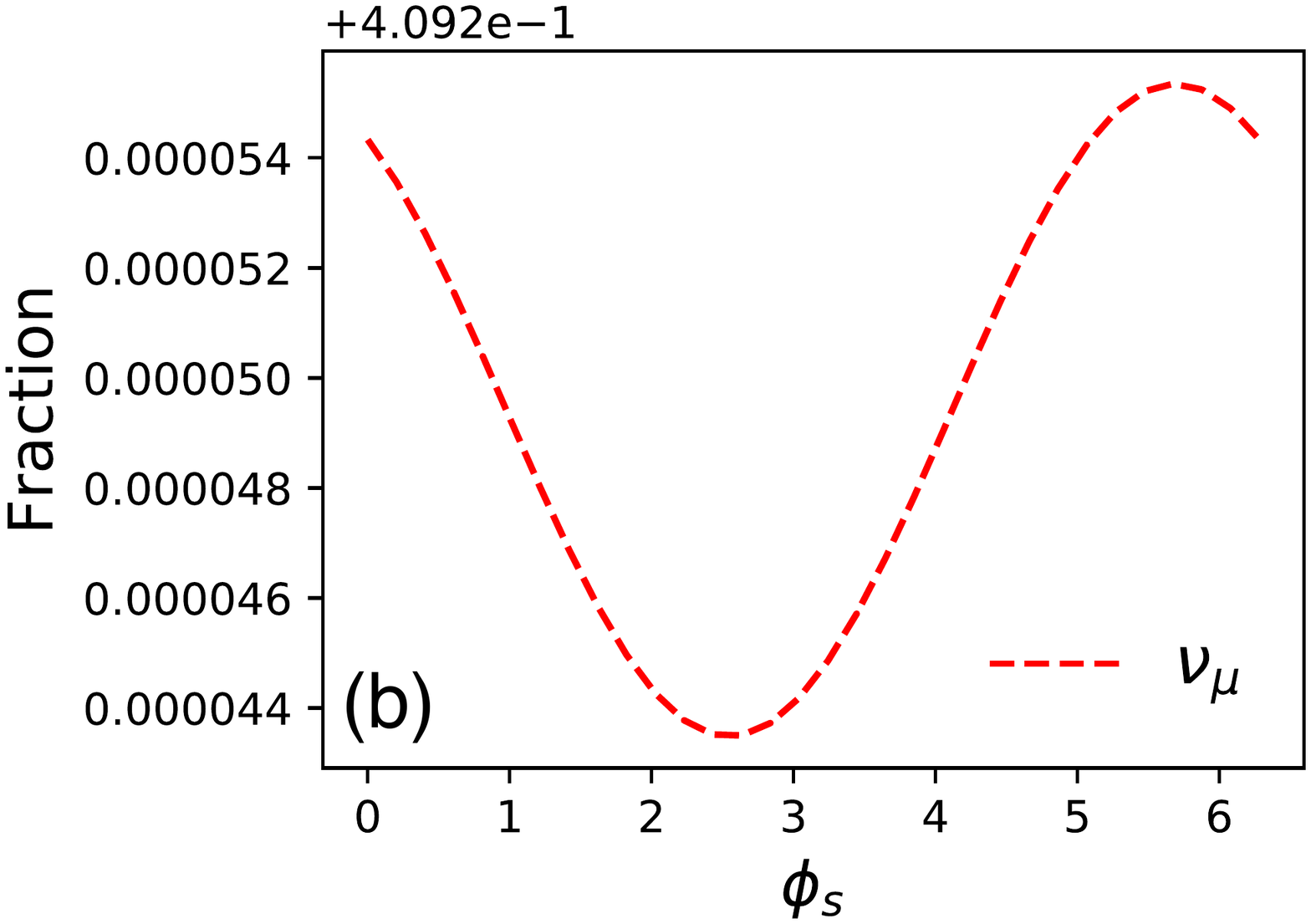}
\includegraphics[trim={0 7.5cm 0 7.5cm},clip,width=0.47\textwidth]{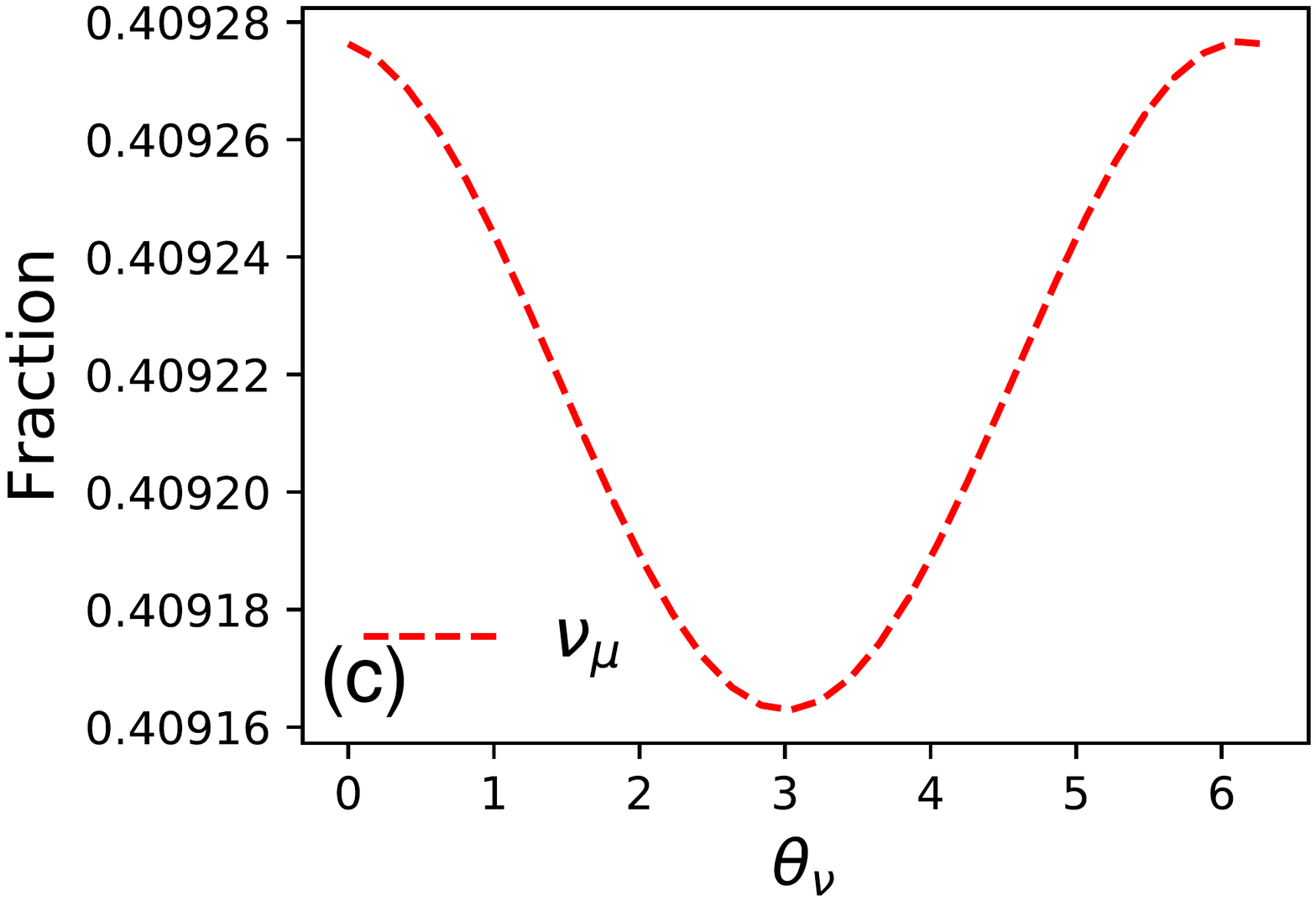}
\includegraphics[trim={0 7.5cm 0 7.5cm},clip,width=0.47\textwidth]{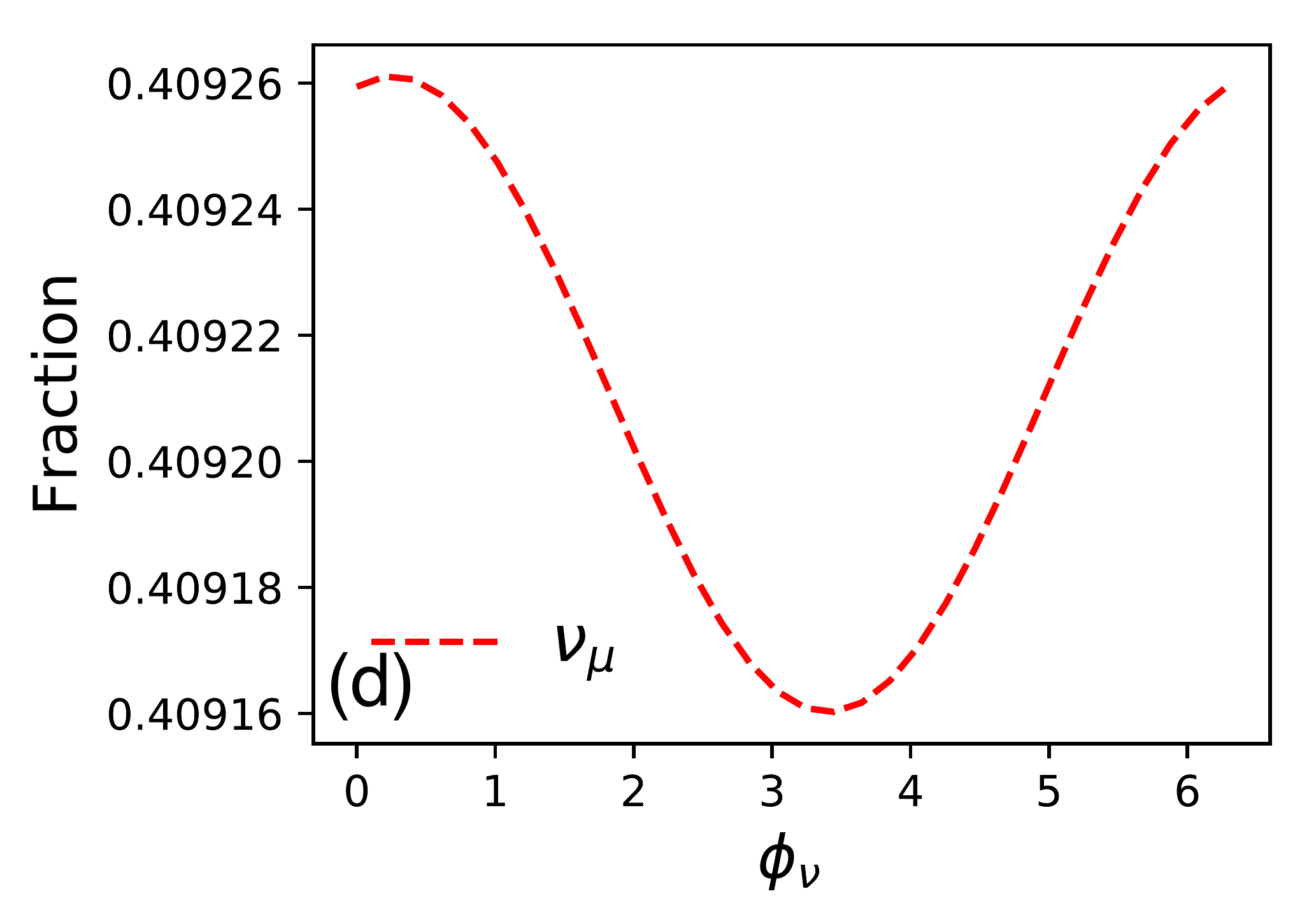}
\caption{The effect of the directional dependence on the flavor
 composition as a function of incoming azimuthal angle (a, c), polar
 angle (d) and the angle corresponding to the seasonal position of Earth
 (b). 
 The DE-induced parameters
 are set to $\theta_{12}, \theta_{13}, \theta_{23} = 0.25 \pi$ and
 $m_{\text{eff}_{21}} = \frac{1}{2}m_{\text{eff}_{31}} = 10^{-26}
 \text{GeV}$, and the flavor composition at the source is set to
 1:2:0. The results are shown for a neutrino energy of $E_{\nu} =
 10^{5}$ GeV, which lies inside the transition range from vacuum
 domination towards DE-induced domination. For energies outside the
 transition phase, the effect is an order of magnitude smaller.}
\label{fig:dirfractions}
\end{figure*}

In Fig.~\ref{fig:dirfractions}, the effect on the final flavor
composition as a function of the different variable angles is shown. 
In Fig.~\ref{fig:dirfractions}a, it is seen that the effect is much
smaller than the effects of the other new parameters explored earlier in
this paper. 
We have to zoom in on one particular flavor to be able to visualize the
effect. 
In Fig.~\ref{fig:dirfractions}b, the fraction of $\nu_{\mu}$ is plotted as
a function of the seasonal shift $\phi_s$. 
The effect is extremely small, with a maximal change of $\sim$0.002\% depending on the season in which the neutrinos are detected. 
Clearly, at this moment, it is far beyond our current abilities to
measure such small differences. 
The advantage of the seasonal shift is that, in the case that a source
produces neutrinos on a regular basis such as blazars, the
flavor ratio of neutrinos originating from that source could be
evaluated in different seasons to search for a seasonal effect. 
The effect of the propagation direction of the neutrinos is slightly
larger than the seasonal effect, as can be seen from
Figs.~\ref{fig:dirfractions}c and \ref{fig:dirfractions}d, in which the
fraction of $\mu_\nu$ is plotted as a function of the incoming
directions $\theta_\nu$ and $\phi_\nu$ respectively. 
The maximal change between the flavor fractions depending on the
incoming direction is $\sim 0.03\%$. 
Although this effect is an order of magnitude larger than the effect of
the seasonal shift, it is still outside our observational reach to
detect such small effects. 
However, eventually future experiments might become more sensitive, and
meanwhile in the years or decades to come, data are being collected,
contributing to better statistics. 
In the case that new physics with the effects described in
Sec.~\ref{sec:behaviour} is found, the directional effect would be the
evidence for a DE-neutrino coupling rather than some other solution. 
It would be also evidence for a non-cosmological constant type of DE.

\section{Conclusion}\label{sec:concl}

We are only at the beginning stage of collecting data from high energy
neutrinos, and exciting times lie ahead. 
It will not take long before IceCube and KM3NeT will determine if the
measured flavor ratio at Earth is compatible with normal physics. 
We explore a possible origin for new physics results in neutrino
telescopes and how this would establish in measurements here on Earth. 

The physics we investigate is a possible coupling between dark energy
(DE) and neutrinos, which engenders an additional source for neutrino
mixing. 
Such a coupling might exist in the case that DE is a dynamical field
rather than a cosmological constant. 
We study the impact on neutrino oscillations in the three-neutrino
framework and find that this could result in significant observable
effects on Earth. 
The part of the oscillation probability that is induced by DE is
independent of energy in the propagation Hamiltonian, has different sign
for neutrinos and anti-neutrinos, and contains a directional component. 
Furthermore, the probability depends on three extra mixing angles, one
new CP-violating phase, and two independent mass parameters
$m_{\text{eff}}$. 
Because of the energy independency of the DE induced part of the
Hamiltonian, while the vacuum oscillation term is proportional to
$\propto \frac{\Delta m^2}{2E}$, the effect of the DE-neutrino coupling
becomes larger for higher neutrino energies. 
Below are our main findings.

\begin{enumerate}
 \item 
The transition from the energy scale in which vacuum oscillation
dominates, to the energy scale where the DE-induced mixing dominates,
happens around $E_{\nu}m_{\text{eff}} \sim 10^{-20}~ \text{GeV}^2$. 

\item
We explored the effect of the coupling on the flavor composition of
astrophysical neutrinos that we would measure on Earth. 
Depending on the flavor composition at the source and the values of the
new mixing parameters, the possible final flavor ratios cover the entire
flavor composition triangle, while vacuum oscillation covers only a
limited area. 
If no tau-neutrinos are produced in astrophysical sources, however, part
of the flavor composition triangle cannot be reached even through mixing
induced by DE.

\item
We also explored the effect on the flavor composition due to the sign
difference in the probability between the neutrinos and
anti-neutrinos and the new CP-violating phase. 
Neutrinos and antineutrinos behave differently over a wide energy range,
which might be possible to detect if the effective mass parameter
$m_{\rm eff}$ happens to have a value between $\sim$10$^{-29}$ and
$\sim$10$^{-25}$ GeV.
In that case, the energy range showing CP-violation covers the Glashow
resonance of $6.3$~PeV, which enables experiments like IceCube and
KM3NeT to distinguish between $\nu_e$ and $\bar{\nu}_e$.

\item
We also determined the sensitivity for current and future experiments to
probe the value of the effective mass parameters $m_{\text{eff}}$. 
We find that current experiments are able to measure anomalous effects
due to the DE-neutrino coupling, and can probe the values of the
new mixing parameters, for a genuine value of the effective mass
parameter down to $m_{\text{eff}}\sim 10^{-27}$, depending on the number
of detected neutrino events. 
Experiments capable of detecting ultrahigh-energy neutrinos could probe
further down to $m_{\text{eff}}\sim 10^{-30}$. 

\item
Because the cosmic expansion has a preferred frame, namely the rest
frame of the CMB, the value of $m_{\text{eff}}$ does depend on our
velocity with respect to the CMB rest frame, and gets slightly
altered for different propagation directions of the incoming
neutrinos (directional dependence), as well as the position of the
Earth with respect to the Sun (the seasonal dependence). 
These effects are small, resulting only in differences between the
flavor composition on a sub-percentage level, but of big importance
in the case new physics is found, since this effect is an unique
feature of the DE-neutrino coupling.

\end{enumerate}

\acknowledgments
We thank Marc Kamionkowski for his insightful comments at an early phase
of this project.
This work is part of the research programme of the Foundation for
Fundamental Research on Matter (FOM), which is part of the Netherlands
Organisation for Scientific Research (NWO) (N.K. and S.A.) and partly
financed by the Netherlands Organisation for Scientific Research (NWO)
through a Vidi grant and JSPS KAKENHI Grant Number JP17H04836 (S.A.).

\appendix

\section{The amplitude of the flavour transition}
\label{app:Amplitude of flavour transition}

The exponential of an $N\times N$ matrix $M$ can be expressed as
\begin{equation}\label{eq:first}
a_0I+a_1M+....+a_{N-1}M^{N-1} .
\end{equation}
The matrix $M$ can also be expressed as
\begin{equation}\label{eq:second}
M = M_0 + \frac{1}{N}(\text{tr}M)I,
\end{equation}
where $M_0$ is an $N \times N$ traceless matrix.
By combining equations \ref{eq:first} and \ref{eq:second}, and defining
the complex phase $\phi \equiv e^{-iL\text{tr}\mathscr{H}_f/3}$ and the
traceless matrix $T\equiv \mathscr{H}_f-(\text{tr}\mathscr{H}_f)I/3$,
we can write
\begin{equation}
e^{-i\mathscr{H}_f L}= \phi e^{-iLT} = \phi(a_0I-iLTa_1-L^2T^2a_2).
\end{equation}
The coefficients $a_0, a_1$ and $a_2$ can be computed from the following
system of linear equations:
\begin{eqnarray}
e^{-iL\lambda_1} &=&a_0 - iL\lambda_1 a_1-L^2\lambda_1^2 a_2,\\
e^{-iL\lambda_2} &=&a_0 - iL\lambda_2 a_1-L^2\lambda_2^2 a_2,\\
e^{-iL\lambda_3} &=&a_0 - iL\lambda_3 a_1-L^2\lambda_3^2 a_2,
\end{eqnarray}
where $\lambda_1, \lambda_2$ and $\lambda_3$ are the eigenvalues of $T$,
by solving 
\begin{equation}
\mathbf{a}= \Lambda^{-1}\mathbf{e},
\end{equation}
where
\begin{widetext}
\begin{equation}
\mathbf{e} = \begin{pmatrix} e^{-iL\lambda_1} \\ e^{-iL\lambda_2} \\ e^{-iL\lambda_3} \end{pmatrix}, \,
\Lambda = \begin{pmatrix} 1-iL\lambda_1 - L^2\lambda_1^2 \\ 1-iL\lambda_2 - L^2\lambda_2^2 \\ 1-iL\lambda_3 - L^2\lambda_3^2 \end{pmatrix}, \,
\mathbf{a} = \begin{pmatrix} a_0 \\ a_1\\ a_2 \end{pmatrix},
\end{equation}
such that eventually we have
\begin{eqnarray}
U_f(L)\equiv e^{-i\mathscr{H}_f L} &=& \frac{1}{(\lambda_1-\lambda_2)(\lambda_1-\lambda_3)}\phi e^{-iL\lambda_1}[\lambda_2\lambda_3 I -(\lambda_2+\lambda_3)T+T^2]\nonumber\\
&&{}+\frac{1}{(\lambda_2-\lambda_1)(\lambda_2-\lambda_3)}\phi e^{-iL\lambda_2}[\lambda_1\lambda_3 I -(\lambda_1+\lambda_3)T+T^2]\nonumber\\
&&{}+\frac{1}{(\lambda_3-\lambda_1)(\lambda_3-\lambda_2)}\phi e^{-iL\lambda_3}[\lambda_1\lambda_2 I -(\lambda_1+\lambda_2)T+T^2].
\end{eqnarray}
\end{widetext}
The eigenvalues $\lambda_i$ of $T$ are solutions of the equation
\begin{equation}
\lambda^3 + c_2\lambda^2+c_1\lambda+c_0 = 0,
\end{equation}
where 
\begin{eqnarray}
c_0 &=&-\text{det}T, \\
c_1 &=& T_{11}T_{22}-T_{12}T_{21}+T_{11}T_{33}-T_{13}T_{31}\nonumber\\
&&{}+T_{22}T_{33}-T_{23}T_{32},\\
c_2 &=&-\text{tr}T.
\end{eqnarray}

\bibliographystyle{h-physrev}
\bibliography{mybib}

\end{document}